\documentclass[prc,nofootinbib]{revtex4}
\usepackage{amssymb}
\usepackage{amsmath}
\usepackage[dvips]{graphicx}
\usepackage{epsfig}
\usepackage{mathrsfs}

\def\d{{\rm d}}

\newcommand{\e}{\textrm{e}}

\newcommand{\p}{{\rm p}}

\begin{document}

\title{Crossover transition in
bag-like models} 

\author{L.~Ferroni, V.~Koch}\affiliation{Nuclear Science Division, Lawrence Berkeley National Laboratory,
1 Cyclotron Road, Berkeley, 94720}

\begin{abstract}
We formulate a simple model for a gas of extended hadrons at zero chemical potential by 
taking inspiration from the compressible bag model.
We show that a crossover transition qualitatively similar to lattice QCD can be 
reproduced by such a system by including some appropriate additional dynamics.
Under certain conditions, at high temperature, the system consist of
a finite number of infinitely extended bags, which occupy the entire space. In this situation
the system behaves as an ideal gas of quarks and gluons. 
\end{abstract}

\maketitle

\section{Introduction}
The phase transition of strongly interacting matter has been intensively studied for 
many years. As early as 1960s, before the discovery of QCD, there was 
speculation about a possible new phase of strongly interacting matter, based on studies of 
the thermodynamics of a hadron gas. Particularly, in the Statistical 
Bootstrap Model~\cite{Hagedorn:1972gv},
the asymptotically exponential mass spectrum of hadrons implied 
the existence of a limiting temperature of about $170$~MeV (the Hagedorn temperature) 
above which hadrons cannot exist. 

After the discovery of QCD and in particular asymptotic 
freedom, the discussion focused on the ideas of a Quark Gluon Plasma (QGP), a system of weakly 
interacting quarks and gluons, and a possible transition between a pion gas and a quark 
gluon plasma. 
The physical picture for such a transition was that at the critical temperature the 
additional degrees of 
freedom carried by the quarks and gluons would be released leading to a rapid increase of the 
entropy, energy-density and pressure. This was supported by first 
Lattice QCD (LQCD) calculations~\cite{creutz}. 
Lately, however, the hadron resonance gas experienced a renaissance mainly due 
to its successful description of hadron yields in heavy ion and also 
elementary particle collisions~\cite{BraunMunzinger:1995bp,Cleymans:1996cd,Broniowski:2002nf,Becattini:2005xt,
Andronic:2008gm,Manninen:2008mg,Becattini:1995if,Becattini:1996gy,Becattini:1997rv,Becattini:2001fg}. 
In addition it was realized that the sharp 
rise of the entropy density near the transition temperature, as observed 
in lattice QCD calculations, could be accounted for by a hadron resonance gas as 
well~\cite{Karsch:2003vd,Cheng:2007jq}. 
The success of the hadron resonance gas below the transition temperature $T_c$, however, 
changes the physical interpretation 
for the QCD phase transition. Instead of releasing additional degrees of freedom 
at $T_c$, the system has to reduce the number of active degrees of freedom at the 
transition temperature, because a hadron gas has many more than a Quark Gluon plasma. 
In case of a Hagedorn exponential mass spectrum, the entropy density actually diverges 
at the Hagedorn temperature\footnote{Actually, for given choices of the mass spectrum parameters
the entropy is finite at the Hagedorn temperature, but it diverges at any higher temperature.}. 
But even if  the number of hadrons is restricted to those 
needed for a successful description of hadronic final states in heavy ion and 
elementary particle collisions, the entropy quickly exceeds that observed on the lattice. 

Meanwhile many papers have addressed this issue ranging 
from QCD-based approaches (see for example~\cite{Karsch:2003vd,Rischke:1987xe,Mingmei:2007jx})
to various generalizations of the hadron 
gas~\cite{Hagedorn:1980cv,Rafelski:1980rk,Kapusta:1981ay,Gorenstein:1981fa,Kagiyama:1992kt,
Gorenstein:1998am,Kagiyama:2002rm,Gorenstein:2005rc,Zakout:2006zj,Bugaev:2007ww,Zakout:2007nb}. 
In this article we will 
study the transition region, and we will provide a simple and intuitive 
modification to the hadron gas model in order  to qualitatively 
reproduce the crossover transition observed in Lattice QCD \cite{Aoki:2006we}.
Our calculations are based on the ideas proposed in~\cite{Gorenstein:1998am}, 
where it was shown that under certain  circumstances a gas of extended hadrons could produce phase transitions 
of the first or second order, and also a smooth crossover transition that might be 
qualitatively similar to that of lattice QCD. 

We first observe that for the typical particle density at the transition 
the size of hadrons needs to be taken into account as it leads to a considerable 
suppression of the available phase space. As a result the number of effective degrees of
freedom is reduced. This suppression alone, however, does not explain the ideal gas 
behavior of lattice QCD
at high temperatures, above $T_c$. Additional model assumptions have to be made. 
For example one could explicitly introduce a deconfined phase and then match
the two different phases. For this exercise to work, however, rather detailed assumptions 
about the intrinsic nature of hadrons have
to be made in order to avoid the occurrence of an actual phase transition~\cite{Mingmei:2007jx}. 
In this paper, we follow a different route. 
We adopt the philosophy that the same partition function should describe 
both ``phases''. To this end we need to introduce appropriate dynamics in order to model 
the crossover observed on the lattice. 
This approach is similar in spirit to~\cite{Gorenstein:1998am}. 
We find that the MIT bag model~\cite{Chodos:1974je} of the hadrons is  
well suited for our purposes, because it embodies confined and deconfined 
phases from the very beginning. 
Thus, we will describe hadrons as extended bags of QGP and  we 
will assume an infinite mass spectrum 
of the Hagedorn's type. The additional dynamics needed to describe the transition 
are simply the  elastic interactions between hadrons. They give rise to a  {\em kinetic} 
pressure that in turn ``squeezes'' the bag-like hadrons. We find that the behavior 
of our system depends sensitively on the choice of parameters for the mass spectrum. One can 
obtain either a real phase 
transition ~\cite{Gorenstein:1981fa,Gorenstein:1998am} or, as we shall show, a crossover. 
In addition, in the latter case the specific parametrization for the mass 
spectrum affects the microscopic structure of the gas of  ``compressible'' hadrons at high 
temperatures ($T>T_c$). One finds either a high temperature phase that is populated by 
one or few infinitely large bags, consistent with the usual picture of a QGP. 
Or, for a different choice of parameters, one obtains a system of many, densely packed heavy
hadrons\footnote{In what follows, we will sometimes use the word ``hadron'' 
with its widest meaning without distinguishing among hadronic state such as resonances, 
bags or very short living states such as {\em clusters} (see, for example~\cite{Becattini:2003ft}).}, which 
nonetheless exhibit the thermodynamic properties of a QGP of massless quarks and gluons.

On first sight, our approach appears to be similar to the ideas of percolation 
models~\cite{percth1,Coniglio:1980zz,Fortunato:1999wr,Fortunato:2000ge,Fortunato:2000hg,Fortunato:2000vf}. 
The finite size corrections to the statistical ensemble remove all the configurations with 
overlapping hadrons, resulting in large hadronic states dominating the partition function.
This is similar to percolation. However, there are quite some differences in the specific implementation. 
First, in our model, we do not consider an explicit coupling between the bags as it is done 
in the percolation model of \cite{Fortunato:1999wr,Fortunato:2000ge,Fortunato:2000hg,Fortunato:2000vf}. 
Instead, we take the effect of the kinetic pressure onto 
the bag sizes into account, resulting in a self-consistency relation for the effective bag pressure. 
This is more in the spirit of a mean field description, although we do not introduce an additional 
interaction but simply consider the kinetic pressure.
Second, in contrast to purely geometric percolation, in our model the number of 
hadrons and their sizes are not independent quantities. In a given multihadron state, 
melting two or more hadrons together 
(to form a bigger one) results in a different kinetic pressure and, in turn, in the rearrangement of 
the sizes of all the hadrons. 

Throughout this paper, we will maintain a simple schematic approach to highlight
the main features of the model leaving a more detailed quantitative analysis 
and further generalizations to future work.
We will confine ourselves to the simplest case of nonrelativistic Boltzmann particles and we 
will neglect subtleties such as surface effects and van der Waals type residual interaction among the bags. 

This paper is organized as follows: in Section~\ref{pf} we introduce the main ideas of the model
and derive the grand-canonical partition function for the gas of compressible hadrons. 
In Section~\ref{isobpf} we will use the corresponding isobaric partition function to
perform a comprehensive numerical analysis. We will study the 
pressure, the energy density and the entropy density of the system. We will further analyze 
particles number, the filling fraction and the average mass of particles in the system.

\section{The partition function}
\label{pf}
In this section we will set up the general formalism for our model. Let us start 
with the partition function $Z(V,T)$ of an ideal gas of Boltzmann particles of mass $m$ and
degeneracy $g$ in the nonrelativistic limit. For the subsequent discussion it is advantageous to express the partition function $Z(V,T)$ in a multiplicity expansion, i.e. as a sum of partition functions
$Z_N(V,T)$ for fixed particle numbers $N$:
\begin{equation}
Z(V,T)\equiv \sum_{N=0}^{\infty} Z_N(V,T) \equiv \sum_{N=0}^{\infty} \frac{1}{N!}\left(g V \right)^N \phi(m,T)^N  \; ,
\label{1.1}
\end{equation} 
with
\begin{equation}
\phi(m,T) \equiv \frac{1}{(2 \pi)^3} \int \d^3 \p 
\exp{\left[-\left(\frac{\p^2}{2 m T}+\frac{m}{T}\right)\right]}= \exp{\left[-\frac{m}{T}\right]}
\left(\frac{m T}{2 \pi}\right)^{3/2}\; .
\label{1.2}
\end{equation} 
Here $V$ and $T$ are the volume and the temperature of the system, respectively.
The function $Z$ is the {\em grand-canonical} partition function with vanishing 
chemical potentials. In the context of this paper we shall refer to $Z_N$ as the {\em canonical} 
partition function keeping in mind that this notation deviates from the conventions for 
relativistic hadron gases, where the canonical ensemble has
fixed Abelian charges (such as electric charge, strangeness, baryon number), but 
no constraints on the number of particles.
Eq.~(\ref{1.1}) can be easily generalized to a multispecies gas of particles. 
If we label with $j=1,\ldots,K$ the various particle species we have:
\begin{equation}
Z(V,T)=\prod_{j=1}^K \left[ \sum_{N_j=0}^{\infty} \frac{1}{N_j!}\left(g_j V \right)^{N_j} \phi(m_j,T)^{N_j}\right]
= \exp{\left[V \sum_{j=1}^{K} g_j \phi(m_j,T)\right]}\; .
\label{1.3}
\end{equation} 
In case of $K\rightarrow \infty$, it is convenient to replace 
the discrete index $j$ with a continuous spectrum density $\rho(m)$ so that 
the number of species in the mass interval $[m,m+\d m]$ is given by $\rho(m)\d m$.
Formally, we make the substitution:
\begin{equation}
\sum_{j=1}^{\infty} g_j \phi(m_j,T)  \rightarrow \int \d m \rho(m) \; \phi(m,T)\; .
\label{1.4}
\end{equation} 
By expanding the exponential in Eq.~(\ref{1.3}) the partition function can then be written as:
\begin{equation}
Z(V,T)= \sum_{N=0}^{\infty} \frac{V^N}{N!}
\left[ \prod_{i=1}^N \int_{0}^{\infty} \d m_i \rho(m_i) \; \phi(m_i,T) \right] \; .
\label{1.5}
\end{equation}

Because a hadron gas, or, more precisely, a Hagedorn gas is characterized by an exponential mass spectrum, we 
set
\begin{equation}
\rho(m)=c_0 \frac{\e^{m/T_0}}{m^\alpha} \;,
\label{1.7}
\end{equation}
where the parameters $c_0$ and $\alpha$ will be determined from empirical data. In the case of  
gas-of-bags models, which also have an exponential mass spectrum, the parameters  
$c_0$ and $\alpha$ will have to be determined from the underlying (dynamical) model 
parameters, such as bag pressure, and so on. We note that $c_0$ has dimensions of $[\rm mass]^{\alpha-1}$ and   $\alpha$
typically ranges from $\alpha=0$ to $\alpha \sim 7$ depending on the model.
In Eq.~(\ref{1.7}), $T_0$ simply parametrizes the mass spectrum. In the context of the MIT bag model, 
$T_0$ can be interpreted as the effective ``temperature'' inside the bag, as will be 
discussed in section \ref{rpf}. 
By substituting Eq.~(\ref{1.7}) in Eq.~(\ref{1.5}) 
we arrive at the following partition function for a hadron gas
\begin{eqnarray}
\label{1.7a}
Z(V,T) &=& \sum_{N=0}^{\infty} \left(\frac{T}{2 \pi}\right)^{3N/2} \frac{(V c_0)^N}{N!}
\left[ \prod_{i=1}^N \int_{0}^{\infty} \d m_i \; m_i^{3/2-\alpha}  \right] \\ \nonumber
&&\times 
\exp{\left[\frac{\sum_{i=1}^N m_i}{T_0}-\frac{\sum_{i=1}^N m_i}{T}\right]} \; .
\end{eqnarray}
This partition function $Z(V,T)$ (and also $Z_N(V,T)$) is  
divergent for $T > T_0$ as already pointed out by Hagedorn. Although an 
upper limit in the mass spectrum $\rho(m)$ regulates the divergences, 
it will not prevent the system from having a much higher entropy density than that 
observed on the lattice.

An exponential mass spectrum without any cutoff may certainly be an oversimplification 
and a more realistic calculation may take into account discrete states as well as a mass 
spectrum that grows less than exponential above a certain mass. However, empirically the 
known hadronic states do indeed grow exponentially up to a mass of $m\sim 2\, \rm GeV$. 
Above that, very few states are known and it is not clear if this is an indication of a 
saturating density of states or simply the lack of experimental data on higher mass resonances. 
Therefore, working with an exponential mass spectrum without any cutoff appears to be 
an approximation as good as any other.
Furthermore, because we are interested only in bulk thermodynamic 
quantities such as energy density and pressure, the use of a continuous mass spectrum should 
be a reasonable approximation as all these quantities represent integrals/sums over the 
mass spectrum. Therefore, in  this paper 
we will assume that the mass spectrum is of the Hagedorn type and will 
discuss a dynamical scenario in the framework of the MIT bag model, which will regulate 
the partition function. 


\subsection{The regularized partition function}
\label{rpf}
To develop the partition function of our model, we need to recall some of the basic 
features of the MIT bag model~\cite{Chodos:1974je}. 
In its simplest formulation, hadrons can be considered as 
{\em bags} of partonic fields confined in a spatial region
with a constant potential energy per unit volume $B$, where $B$ is commonly referred 
as the bag constant or bag pressure.
The total energy, i.e. the mass $m$ of a bag with volume $V_b$ is then given by~\cite{Chodos:1974je}:
\begin{equation}
m=U+BV_b \; ,
\label{bag1}
\end{equation} 
where $U$ is the internal energy of the field inside the bag.
When its linear extension is larger than the wavelengths of the
partons (the quanta of the inner field), we can approximate the bag 
by a gas of free massless particles confined to its volume~\cite{Chodos:1974je}. 
For a sufficiently large $V_b$, the internal energy is then given by the relation:
\begin{equation}
U=3 p_r V_b \; ,
\label{pgas}
\end{equation}
where $p_r$ is the pressure of the gas. For a single hadron the stability condition requires:
\begin{equation}
p_r \equiv B \; ,
\label{dum}
\end{equation}
which then gives
\begin{equation}
m=4 B V_b \; .
\label{massbagm}
\end{equation}
The effective temperature of the bag is related to the bag pressure by the relation
$T_0 \equiv kB^{1/4}$, where $k$ is a dimensionless constant whose value
depends on the number of internal degrees of freedom of the gas inside the bag.
For large $V_b$, the entropy $S$ of a bag, is the entropy of a massless
gas with internal energy $U$ and pressure $B$~\cite{Chodos:1974je}, therefore\footnote{In principle, on the
left-hand side of Eq.~(\ref{entrobag}) one should subtract a constant $S_0$ which corresponds 
to the entropy at $U=0$. Here, this term has been omitted as it is immaterial for our purposes.},
\begin{equation}
S= \frac{4U}{3kB^{1/4}} \equiv  \frac{4U}{3T_0}= \frac{m}{T_0} \;.
\label{entrobag}
\end{equation}
From Eq.~(\ref{entrobag}) one can derive the level density 
\begin{equation}
\rho(m) \propto \e^S \propto \e^{m/T_0} \;.
\label{baglevdens}
\end{equation}
Note that the generic spectrum introduced in Eq.~(\ref{1.7}), has an additional 
contribution: $m^{-\alpha}$.
This factor can be interpreted as a logarithmic correction to the entropy of the bag.
In what follows, we will retain the spectrum in Eq.~(\ref{1.7}) and we will analyze 
different values of $\alpha$. Of course, Eq.~(\ref{baglevdens})
will correspond to the case $\alpha=0$. 

Once the temperature $T$ of the gas of bags approaches $T_0$,
the average masses and hence the volume (see Eq.~(\ref{massbagm})) of the bags grow
very fast. Therefore, the bag-like hadrons tend to occupy 
more and more of the available space and, eventually, they will overlap. 
To avoid multiple counting of the phase space, 
configurations with overlapping bags need to be excluded from 
the partition function. For a finite system of volume $V$ this can be achieved with an 
excluded volume correction, where the total volume $V$ is replaced by the available 
volume $(V-\sum_{i=1}^N V_i)^N$, where $V_i$ is the volume of the $i-$th particle. 
In addition the volume of all bags $\sum_{i=1}^N V_i$ 
should not exceed the total volume $V$.  
Following Ref.~\cite{Gorenstein:1981fa}, this leads to the modified
 $N$-particle phase space integral
\begin{equation}
\left[\prod_{i=1}^N \frac{V}{(2 \pi)^3} \int \d^3 \p \right] \rightarrow 
\left[\prod_{i=1}^N \frac{1}{(2 \pi)^3} \int \d^3 \p \right] \left(V-\sum_{i=1}^N V_i\right)^N
\Theta{\left( V-\sum_{i=1}^N V_i \right)} \; .
\label{nphsp}
\end{equation} 
This modified phase space integral results in a well-defined and finite  partition function 
at every temperature
\begin{eqnarray}
\label{1.7b}
&&Z(V,T) = \sum_{N=0}^{\infty} \left(\frac{T}{2 \pi}\right)^{3N/2} \frac{c_0^N}{N!}
\left[ \prod_{i=1}^N \int_{0}^{\infty} \d m_i \; m_i^{3/2-\alpha}  \right] \\ \nonumber
&&\times 
\exp{\left[\frac{\sum_{i=1}^N m_i}{T_0}-\frac{\sum_{i=1}^N m_i}{T}\right]}\left(V-\sum_{i=1}^N V_i\right)^N
\Theta{\left( V-\sum_{i=1}^N V_i \right)} \; .
\end{eqnarray}
It can be shown that such a system of extended hadrons 
leads to a constant value for the energy-density $\varepsilon$, in contradiction with Lattice QCD, where the energy 
density is found to increase with the fourth power of the temperature, $\epsilon_{\rm Lattice}\sim T^4$. In fact, 
for $T\rightarrow \infty$,
the most favorite configurations are those where the hadrons occupy all the available space. In this case,
the energy density of the system correspond to $\varepsilon=4 B$, i.e. the inner density of the hadrons 
(see Eq.~(\ref{massbagm})). The underlying reason for this behavior is that the system is not able to pick 
up additional kinetic energy once the entire volume is filled with bags, as the bags have no more room to move.

Obviously some additional dynamics needs to be included to allow for the system to pick up more 
energy as the temperature is increased.
To this end we adopt the idea of compressible bags~\cite{Kagiyama:2002rm}.
More precisely, we will allow the volume of the hadrons to vary under the 
effect of the pressure generated by their own thermal motion in a self-consistent way. 
Consequently, as the temperature and hence the pressure increase, the bags will be compressed and 
acquire a higher internal mass/energy density. We will show that the system does not  
exhibit any limiting value of the energy density, and, under appropriate conditions, 
exhibits the desired increase of the energy density and entropy.

To illustrate the underlying mechanism, 
let us consider a gas of many hadrons. For small temperatures,  $T \ll T_0$, the system 
is dilute ($V \gg \sum_{i=1} V_i$) and behaves like a gas of noninteracting point-particles. 
With increasing temperature, the average mass, and hence the spatial extent of the hadrons, 
increases and as $T$ approaches $T_0$ the dilute-gas approximation seizes to be valid. 
The pressure exerted by the other particles becomes sizable and its effect on the hadrons 
properties, such as the size, can no longer be ignored. In other words, 
in addition to the bag pressure $B$, every particle in the system will feel an additional \emph{kinetic} pressure
$p_k$ that is generated by the thermal motion of the other hadrons in the gas.
In this situation, the stability condition,  Eq.~(\ref{dum}), needs to be modified by taking 
onto account the contribution of the kinetic pressure $p_k$.
Microscopically, the pressure $p_k$ can be interpreted as the 
consequence of elastic collisions. The effect of inelastic collisions, which are certainly 
present in a hadron gas, in our approach are accounted by the infinite mass spectrum of hadrons
without enforcing any constraint on the number of particles $N$.
In this way all 
the possible configurations with few large hadrons or many small ones are included.
This is analogous to the hadron-resonance gas model, 
where a large part of the inelastic interaction is taken into account by adding resonances 
as free particles in the gas.

\begin{figure}[h!]
\begin{center}
\includegraphics[width=0.7\textwidth]{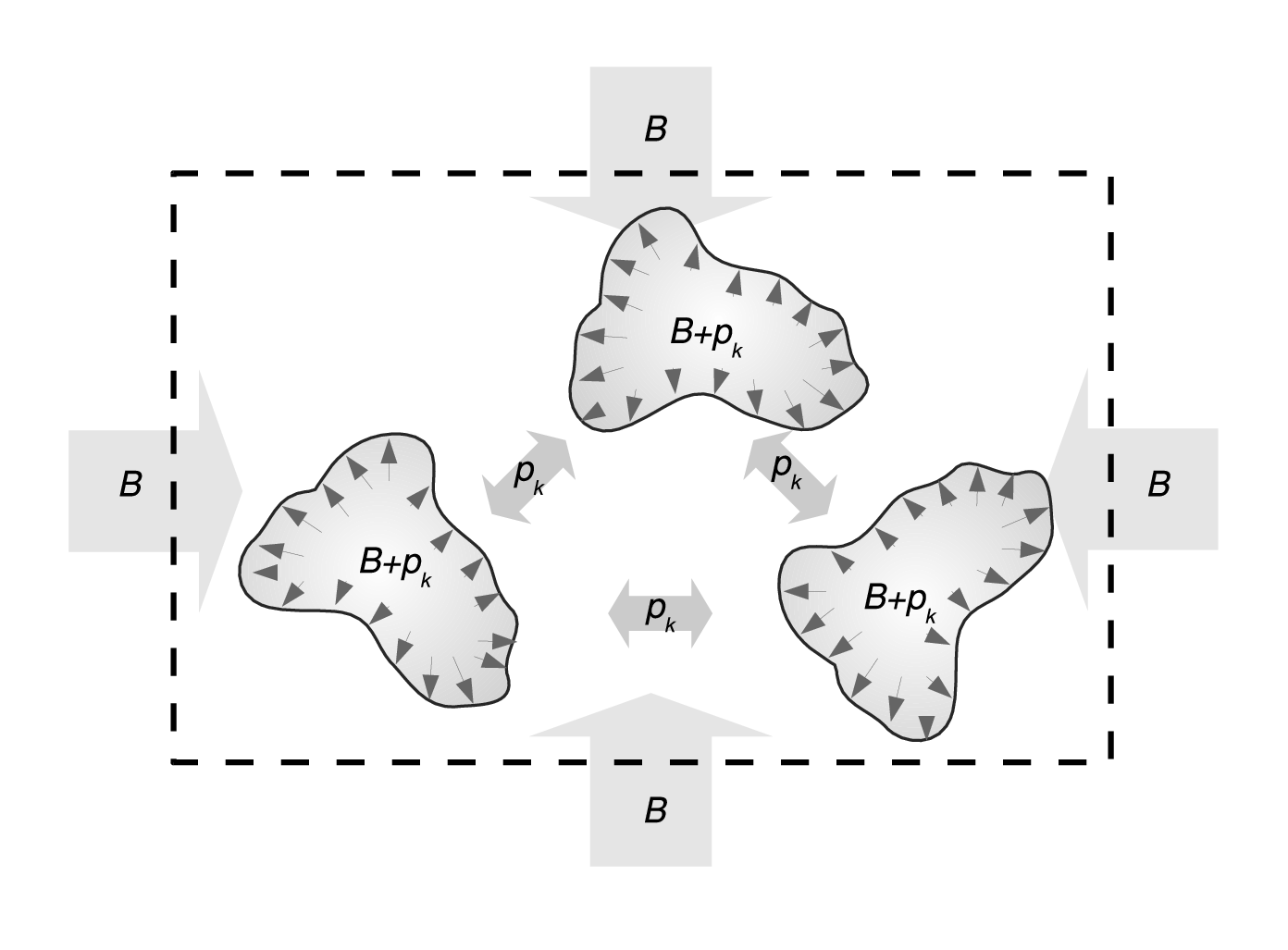}
\caption{\small{Pictorial representation of the pressure balance in a gas of 
compressible hadrons. The internal hadron pressure must be equal to the sum of the constant bag 
pressure $B$ plus the pressure $p_k$ generated by the thermal motion of the hadrons themselves.}}
\label{bp}
\end{center}
\end{figure}

Neglecting any surface effect, 
the simplest generalization of the stability condition, Eq.~(\ref{dum}), is 
\begin{equation}
p_r=B+p_k(V,T) \; .
\label{simplass}
\end{equation}  
A pictorial illustration of the pressure balance in the last equation is given in Fig.~(\ref{bp}).
Each hadron in the gas is characterized by the same 
internal pressure $p_r$. But instead
of being a constant (as in the case of a single hadron in the vacuum where $p_r \equiv B$), 
$p_r$ now depends on $T$ and $V$, and must 
be evaluated in a self-consistent fashion from the partition function itself.

It is clear that the number of hadrons, their sizes and the kinetic pressure are 
all connected by the above self-consistency relation. Accordingly, as we have already pointed out, if we split 
or combine two or more hadrons, the pressure, and thus the volume 
of all the hadrons in the system, changes.
This new state corresponds to a distinct
state of the ensemble that cannot be obtained by a simple geometrical clustering procedure as usually
done in percolation models~\cite{percth1,Coniglio:1980zz,Fortunato:1999wr,Fortunato:2000ge,Fortunato:2000hg,Fortunato:2000vf}.

To account for this additional dynamics, we need to generalize the 
partition function, Eq.~(\ref{1.7b}).
To this end, we  write the explicit dependence of volumes and the masses of the bags on the pressure $p_r$
 \begin{eqnarray}
\label{replacements}
V_i &=& \frac{U_i}{3 p_r} \\ \nonumber
m_i &=&  U_i + B V_i = U_i \left(1+\frac{B}{3 p_r} \right) \; .
\end{eqnarray}
We also rewrite the exponential mass spectrum, $\sim e^{m_i/T_0}$ in terms of the general expression for the entropy
\begin{equation}
S_i=\frac{4 U_i}{3 k p_r^{1/4}} \; ,
\label{entropy}
\end{equation}
leading to the substitution
\begin{equation}
\e^{m_i/T_0}\qquad \rightarrow \qquad\e^{4 U_i/3 k p_r^{1/4}}
\end{equation}
Finally, because the bag masses depend on $p_r$, instead 
of integrating on $\d m_1\ldots \d m_N$ as in Eq.~(\ref{1.7b}) we will perform 
the integral over the internal energies, i.e. we substitute 
\begin{equation}
\label{intvar}
 \int \d m_i \qquad \rightarrow \qquad \frac{4}{3} \int
\d U_i  \; .
\end{equation}
The factor $4/3$ in the previous formula ensures that we recover the partition function, Eq.~(\ref{1.7b}) 
in the limit of $p_k\rightarrow 0$, i.e. in the dilute gas limit.
With the replacements in Eq.~(\ref{replacements}) to (\ref{intvar}), the modified grand-canonical 
partition function can now be written on the 
basis of Eq.~(\ref{1.7b}) and reads:
\begin{eqnarray}
\label{gcpartfunUpr}
&&Z(V,T) = \sum_{N=0}^{\infty} \left( \frac{4}{3} \right)^N
\left(\frac{T}{2 \pi}\right)^{3N/2}\frac{c_0^{N}}{N!}
\\ \nonumber \; 
&&\times\left[ \prod_{i=1}^N \int \d U_i 
\left(U_i+ B \frac{U_i}{3p_r} \right)^{3/2-\alpha} 
\right] \exp{\left\{\left[ \frac{4}{3kp_r^{1/4}}-\frac{1}{T}\left(1+\frac{B}{3p_r}\right)
\right]\sum_{i=1}^N U_i\right\}}\\ \nonumber \; 
&&\times 
\left(V-\sum_{i=1}^N \frac{U_i}{3p_r} \right)^N \Theta{\left( V-\sum_{i=1}^N\frac{U_i}{3p_r}  
\right)}\; .
\end{eqnarray}
The new partition function $Z(V,T)$ is identical (by construction) 
to Eq.~(\ref{1.7b}) when $p_r \equiv B$. 
Conversely, for a given set $\{U_i\}$, the effect of a finite kinetic pressure $p_k > 0$  
is to squeeze each bag to a smaller size (see Eq.~(\ref{replacements})), and  as a result,
the effective bag temperature $T_b \equiv k p_r^{1/4}$ increases.
The Eq.~(\ref{gcpartfunUpr}) can be made more familiar by substituting
\begin{equation}
\label{newvar}
U_i\rightarrow \eta_i=4 U_i / 3
\end{equation}
leading to
\begin{eqnarray}
\label{partfunmpr}
Z(V,T) &=& \sum_{N=0}^{\infty} \left(\frac{T}{2 \pi}\right)^{3N/2}\frac{c_0^{N}}{N!}
\left[ \prod_{i=1}^N \int \d \eta_i 
\left(\frac{3}{4} \eta_i+ B \frac{\eta_i}{4p_r} \right)^{3/2-\alpha} 
\right]\\ \nonumber \; 
&\times& 
\exp{\left\{\left[ \frac{1}{kp_r^{1/4}}-\frac{1}{T}\left(\frac{3}{4}+\frac{B}{4p_r}\right)
\right]\sum_{i=1}^N \eta_i\right\}} \\ \nonumber \; 
&&\times 
\left(V-\sum_{i=1}^N \frac{\eta_i}{4p_r} \right)^N \Theta{\left( V-\sum_{i=1}^N\frac{\eta_i}{4p_r}  
\right)}\; .
\end{eqnarray}
Obviously, in the dilute gas limit, $p_k\rightarrow 0$  $\eta_i\rightarrow m_i$. 

We further introduce a lower bound $m_c$ for the integrals  over 
$\{ \d \eta_i \}$. This is needed because, for $\alpha>0$, the spectrum in 
Eq.~(\ref{1.7}) has a pole in $m = 0$, resulting in a divergent partition function  
for $\alpha \geq 5/2$. 
Because there are no hadrons lighter than the pion, 
we will set $m_c\equiv m_{\pi}=0.139$~GeV.

\section{The isobaric partition function}
\label{isobpf}

\begin{figure}[h!]
\begin{center}
\includegraphics[width=0.8\textwidth]{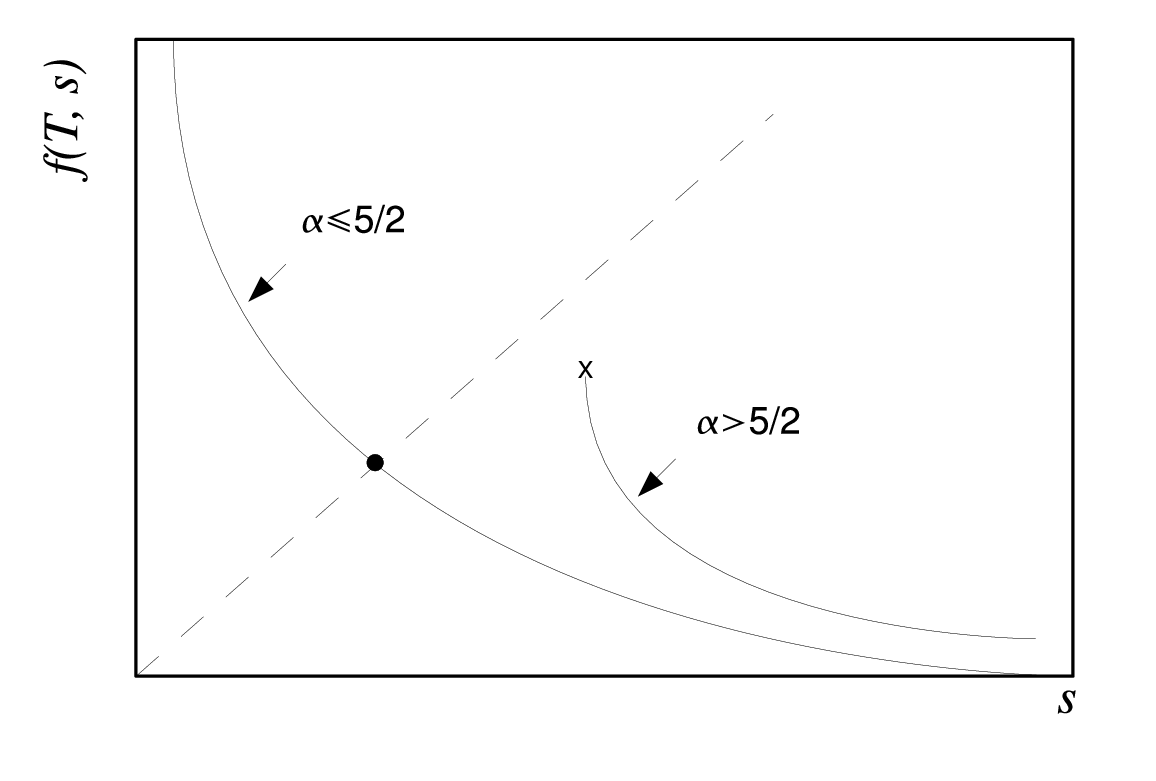}
\caption{\small{Schematic representation of two possible curves $f(T,s)$. When $\alpha \leq 5/2$ the solution 
of Eq.~(\ref{ibpf4}) always corresponds to the rightmost singularity (denoted with the black dot), whereas for 
$\alpha > 5/2$ the function $f(T,s_f(T))$ is finite, and when $T>T_0$ it can happen that 
$s_f(T)$ is the only singularity (denoted with the X). 
The figure is adapted from~\cite{Gorenstein:1981fa,Gorenstein:1998am}.}}
\label{fts}
\end{center}
\end{figure}

Because the pressure $p_k$ is thermally generated, it must be calculated 
from the partition function itself, resulting in a self-consistency relation.
This is best achieved by introducing the isobaric partition function, which 
is defined as the Laplace transform of $Z(V,T)$ over the variable $V$:
\begin{equation}
\widehat{Z}(T,s)\equiv \int_{0}^{\infty} \d V Z(V,T)\exp{\left[-sV\right]} \; .
\label{ibpf1}
\end{equation}
The quantity $s T$ in Eq.~(\ref{ibpf1}) plays the role of a constant external pressure. 
Accordingly, the equilibrium condition requires
\begin{equation}
p_k= s T \; .
\label{pkst}
\end{equation}
The integral in Eq.~(\ref{ibpf1}) can be solved analytically (see Appendix A) and gives:
\begin{equation}
\widehat{Z}(T,s)=\frac{1}{s}\sum_{N=0}^{\infty}\left[
\frac{f(T,s)}{s}\right]^N=\frac{1}{s-f(T,s)}
\label{ibpf2}
\end{equation}
with
\begin{eqnarray}
\label{ibpf3a}
f(T,s)&=&c_0 \left(\frac{T}{2 \pi}\right)^{3/2} \int_{m_c}^{\infty} \d \eta 
\left(\frac{3}{4} \eta+ B\frac{\eta}{4(B+ s T)} \right)^{3/2-\alpha} \\ \nonumber \; 
&\times& \exp{\left[ \frac{\eta}{k(B+sT)^{1/4}} - \frac{\eta}{T} \right]} \; .
\end{eqnarray}
In the limit $V\rightarrow \infty$ the asymptotic behavior of $Z(V,T)$  
is defined by the singularity of $\widehat{Z}(T,s)$ with the largest 
real part~\cite{Gorenstein:1981fa}.
We have two distinct cases: $\alpha \leq 5/2$ and $\alpha > 5/2$.
For $\alpha \leq 5/2$ the transform 
$\widehat{Z}(T,s)$ in Eq.~(\ref{ibpf2}) has two kinds of singularities: the first, $s_0(T)$, 
is given by the pole of $1/(s-f(T,s))$, i.e.
\begin{equation}
s_0(T)=f(T,s_0(T)) \; ,
\label{ibpf4}
\end{equation}
whose solution is the pressure of the system according to Eq.~(\ref{pkst}).
The second singularity, $s_f(T)$, corresponds to a divergence of 
the function $f$ itself. This happens if the exponent of the integrand of $f(T,s)$ in 
Eq.~(\ref{ibpf3a}) vanishes, i.e.   
\begin{equation}
s_f(T) = \frac{T^3}{k^4}-\frac{B}{T} \; .
\label{sing}
\end{equation}
The situation is schematically represented by the leftmost curve 
in Fig.~(\ref{fts}).
The solid line represents 
the function $f(T,s)$ for a given temperature and the $45^\circ$'s dashed line 
corresponds to $s$.
The intersection between the dashed and the solid line corresponds to the solution $s=s_0(T)$ of 
Eq.~(\ref{ibpf4}) and is denoted by a black dot.
The function $f(T,s)$, is a positive function of $s$ that goes to infinity
for $s=s_f(T)$ and tends to zero as $s \rightarrow \infty$.
Consequently, there is always a solution for Eq.~(\ref{ibpf4}) and the 
pole $s=s_0(T)=p_k(\infty,T)/T$ corresponds to the rightmost singularity. 
Thus the pressure has always a solution.
For $\alpha > 5/2$ the situation is different, however.
In this case the function $f$ has an essential discontinuity 
at $s_f(T)$: it is finite at $s=s_f(T)$, and diverges 
for $s < s_f(T)$. Since $s_f(T)$ increases with temperature (see Eq.~(\ref{sing})) and $f(T,s)\rightarrow 0$ as $s \rightarrow \infty$, for sufficiently large $T$, 
$f(T,s_f(T))<s_f(T)$, and, consequently,  Eq.~(\ref{ibpf4}) does not have a solution. This situation is illustrated by the rightmost curve in Fig.~(\ref{fts}) where
 $f(T,s_f(T))$ (denoted by a X) lies below the diagonal\footnote{
Notice that this can only occur for temperatures $T> k B^{1/4} \equiv T_0$ since the singularity $s_f(T)<0$ for $T<T_0$
(see Eq.~(\ref{sing})).}.
Both these cases, have been discussed in~\cite{Gorenstein:1981fa,Gorenstein:1998am}
where the absence of the solution $s_0(T)$ was identified with the onset of a phase transition.
Here, we analyze in detail the case of a crossover transition, i.e. $\alpha \leq 5/2$. 

\begin{figure}[h!]
\begin{center}
\includegraphics[width=0.9\textwidth]{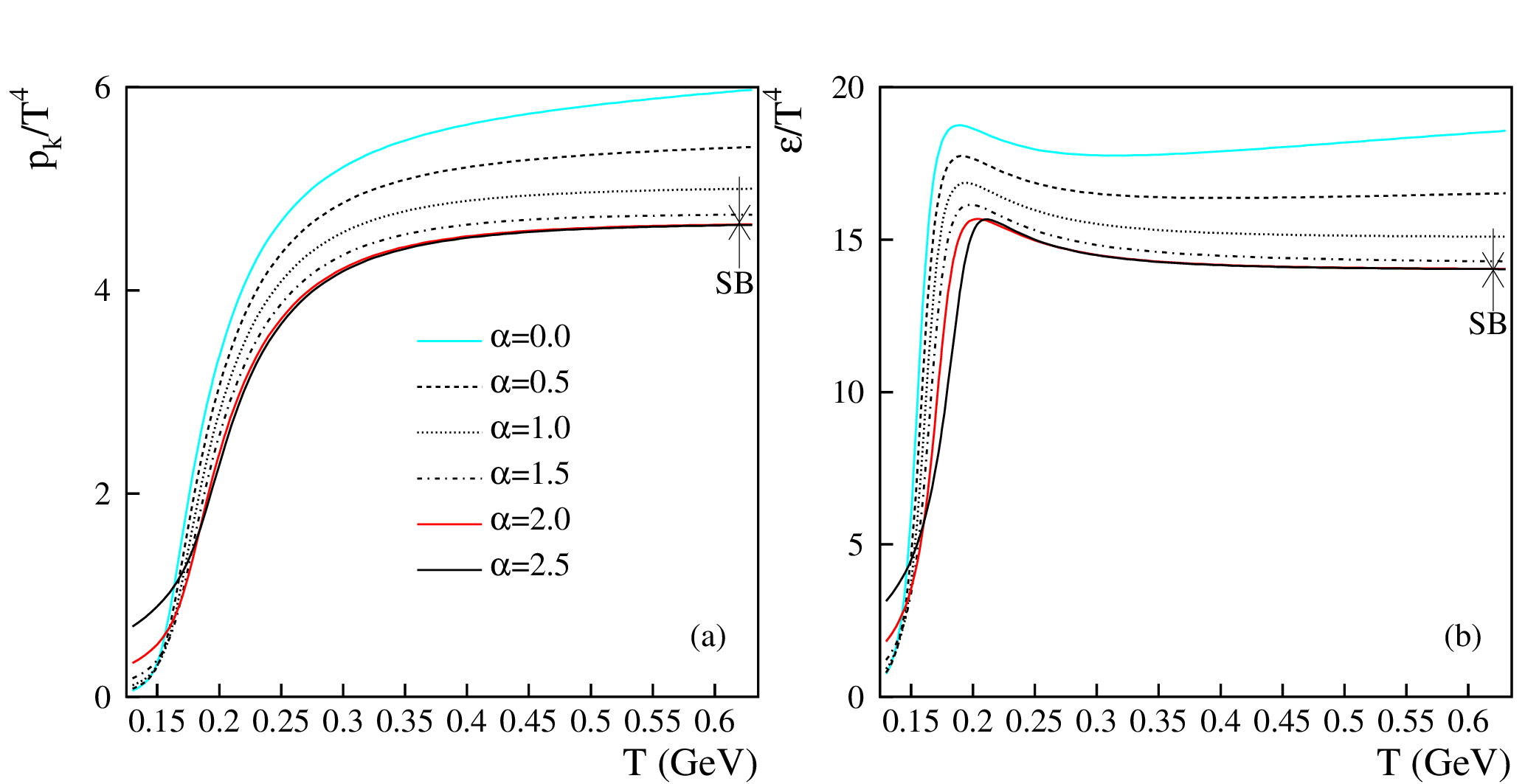}
\caption{\small{(Color online) Left panel (a): The ratio $p_k(\infty,T)/T^4$ calculated with 
the isobaric partition function for various values of $\alpha$. Above $T_0$, 
the curves at $\alpha=2.0$ and $\alpha=2.5$ are practically coincident.
Right panel (b): The corresponding value of $\varepsilon/T^4$.}}
\label{isob1}
\end{center}
\end{figure}

Before we proceed, let us fix the model parameters.
In what follows, we set $k=0.68$ and we keep the product
\begin{equation}
T_0 \equiv kB^{1/4}=0.17  \;{\rm GeV}
\label{3.5}
\end{equation}
fixed. This yields a bag pressure $B=3.9\cdot 10^{-3}$~GeV$^4$ (i.e. $B^{1/4}=250$~MeV) which is 
a plausible value for the bag model parameter.
The constant $k$ has been chosen in order to obtain 
roughly the same value for $\varepsilon/T^4$ as LQCD for large $T$ (see Fig.~(\ref{isob1})).
Its value can also be estimated by counting the degrees of freedom of a thermal system of independent 
quarks and gluons . Our choice lies between the values for the lightest quark doublet u, d 
($k=0.70$) and for u, d and s quark ($k=0.66$).
The  values for the remaining parameter $c_0$ have been fixed by fitting the shape of 
the actual hadron mass spectrum over the mass range of $1\,-\,2\,\rm GeV$. They are given in Table~\ref{tab1}.
\begin{table}[h!]
\begin{center}
\begin{tabular}{|l|c|c|c|c|c|c|}
\hline
$\alpha$ & $0$ & $1/2$ & $1$ & $3/2$ & $2$ & $5/2$ \\
\hline
$c_0$ (GeV$^{\alpha -1}$) & $0.157$ & $0.199$ & $0.252$ & $0.318$ & $0.400$ & $0.502$ \\
\hline
\end{tabular}
\end{center} 
\caption{\small{Standard values of the parameter $c_0$ for different choices of $\alpha$.}}
\label{tab1}
\end{table}
Of course a different mass range or a different choice for $T_0$ would affect these fits. 
However, for our schematic considerations here, 
a fine tuning of the model parameters is rather meaningless. 
In the same spirit
we also ignore the $\sim 10 \%$ deviation of the LQCD result for $\varepsilon/T^4$ 
from the free gas (Stefan-Boltzmann) limit~\cite{Cheng:2007jq}.

In Fig.~(\ref{isob1}a) we plot the pressure $p_k(\infty, T)/T^4 \equiv s_0(T)/T^3$ 
evaluated with the isobaric partition function, Eq.~(\ref{ibpf4}).
As one can see, the results depend on the choice of $\alpha$.
For $\alpha=0$ and $1/2$, the curves grow with the temperature
with larger slopes for smaller $\alpha$'s.
Instead, for $1 \leq \alpha \leq 5/2$, they settle onto constant asymptotic values 
(as we have verified numerically up to $T \sim 6$~GeV).
As $\alpha$ change from $1$ to $5/2$ the asymptotic value converges very fast to 
$1/k^4=4.67$ (the  
Stefan-Boltzmann limit\footnote{In the situation where the system is completely filled by the 
inner hadrons matter (i.e. the free
massless gas), the relation between pressure and temperature is $p=(T/k)^4$. 
Accordingly, the energy density is given
by $\varepsilon = 3p = 3(T/k)^4$.}) from above. As shown in the plot, the curves practically coincide with 
the Stefan-Boltzmann limit already for $\alpha=2$.
This behavior can be understood by inspecting the solution $s_0(T)=p_k(\infty,T)/T$ 
for the pressure. Because $s_0(T)$ is always larger than $s_f(T)$, we have
\begin{equation}
p_k(\infty,T) > \frac{T^4}{k^4}-B \; 
\label{presslim}
\end{equation}
and for  $T \rightarrow \infty$
\begin{equation}
\frac{p_k(\infty,T)}{T^4} \gtrsim \frac{1}{k^4} \; .
\label{presslim2}
\end{equation}
The pressure $p_k(\infty,T)$ converges to $T^4/k^4$ only for sufficiently large values of $\alpha$, 
when the solutions $s_0(T)$ and $s_f(T)$ get closer and closer as 
$T \rightarrow \infty$\footnote{Actually, since $p_k=s_0(T) T$, to obtain the pressure of an ideal gas 
the difference $(s_0(T)-s_f(T))$ must decrease faster than $1/T$. For the ratio $p_k/T^4$ it is sufficient that 
the difference $(s_0(T)-s_f(T))$ grows slower than $T^3$.}.
     
In Fig.~(\ref{isob1}b), we plot the ratio $\varepsilon/T^4$, 
where $\varepsilon$ has been 
evaluated numerically 
by using the relation:
\begin{equation}
\varepsilon=T \frac{\partial p_k}{\partial T}-p_k \; .
\label{isobeps}
\end{equation}
Again this quantity converges to a finite asymptotic limit only for 
$1 \leq \alpha \leq 5/2$ and, as before, coincides with the massless gas 
limit $3/k^4 \sim 14$ for $\alpha=2$ and $\alpha=5/2$.
The overall behavior is roughly the same as LQCD except for a small ``horn'' right above
$T_0$. A closer look reveals that this is due to the constant bag pressure $B$ that produces 
a contribution $\sim B$ to the energy density of the system. 
Being a constant term its contribution to $\epsilon/T^4$ becomes negligible 
at high temperatures.

\begin{figure}[h!]
\begin{center}
\includegraphics[width=0.9\textwidth]{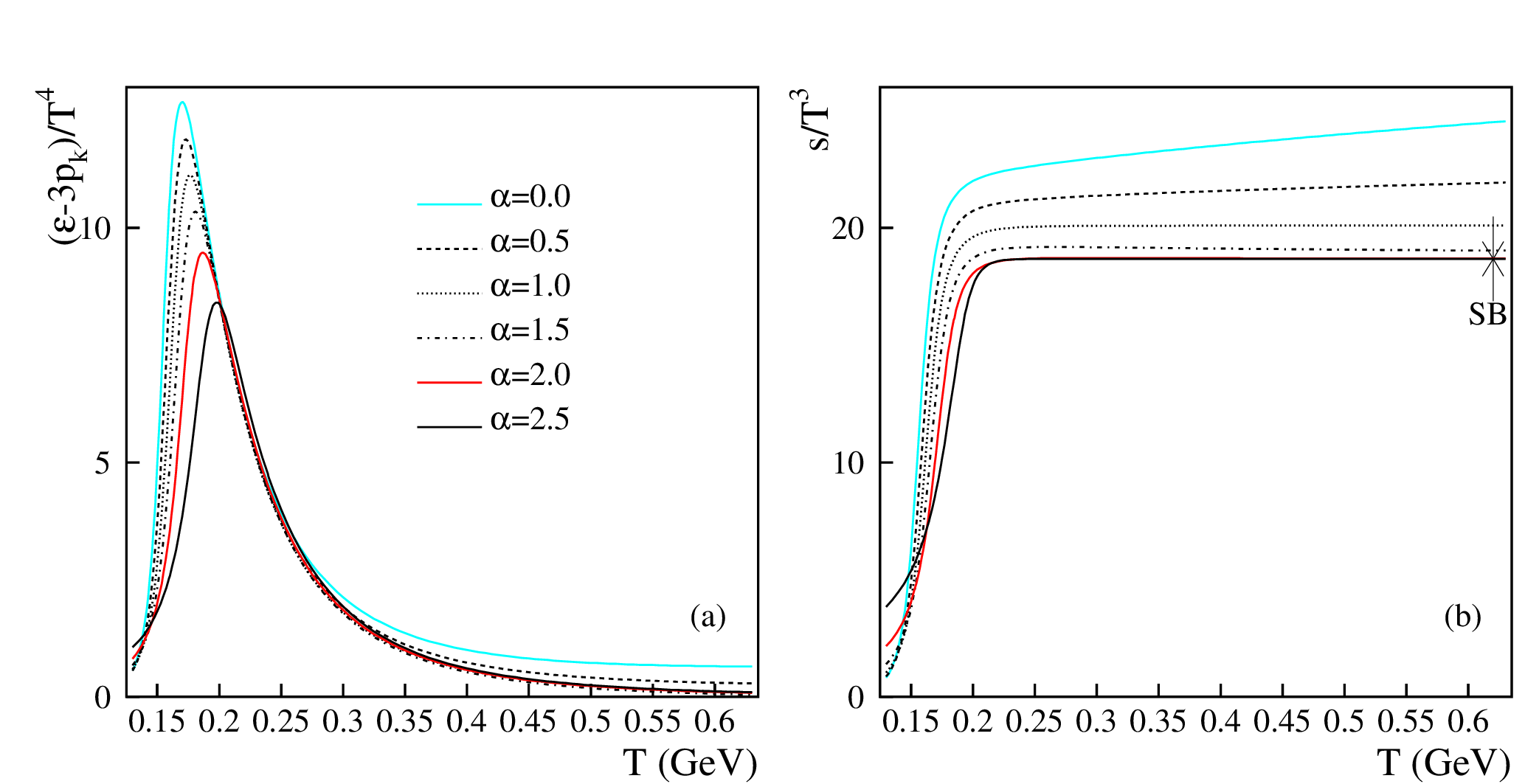}
\caption{\small{(Color online) Left panel (a): the ratio $(\varepsilon-3p_k)/T^4$.
Right panel (b): The ratio $s/T^3$, where $s$ is the entropy-density. }}
\label{isob2}
\end{center}
\end{figure}

In Fig.~(\ref{isob2}a), the ratio $(\varepsilon-3p_k)/T^4$ is plotted up
to $T=0.6$~GeV. This quantity corresponds to the trace of the energy-momentum
tensor $\Theta^{\mu\mu}(T)/T^4$, which is actually the fundamental quantity calculated in LQCD~\cite{Cheng:2007jq}. 
In Fig.~(\ref{isob2}b), we plot the ratio $s/T^3\equiv (\varepsilon+p_k)/T^4$, where
$s$ is the entropy density. 
Again, for $\alpha=0$ and $1/2$, the curves do not converge to a constant value, as expected 
from the previous results for $\varepsilon$ and $p_k$.

In our scheme, the pressure $p_k$ (and thus the energy density and the entropy)  
exceeds the corresponding 
Stefan-Boltzmann limit, except for $\alpha = 2$ and $5/2$ where it converges to it.
To obtain a lower pressure (at least at finite temperatures) one needs to further reduce 
the effective degrees of freedom of the system. This could be possibly achieved 
by introducing a surface energy term. In addition to the fact that part of the energy of the system would be  
spent to create the bags surface, such a contribution would favor spherically 
shaped bags, resulting in a further suppression of the accessible phase space. 
Probably, a similar effect could be also obtained including some residual 
repulsive interaction of the van der Waals type.

We stress that the general behavior of our model for the pressure, energy- and entropy-density cannot 
be obtained by simply introducing an upper 
mass-cutoff on the exponential spectrum in Eq.~(\ref{1.7a}): the entropy-density would considerably 
exceed that obtained from 
LQCD even if we kept only masses up to $2$~GeV. In addition, the system would reach the massless 
gas limit only at temperatures 
much higher than the cutoff itself.
On the other hand, in our model it is absolutely essential to assume an infinite mass spectrum. 
Otherwise the 
flat behavior in Fig.~(\ref{isob1}) and on Fig.~(\ref{isob2}b) would be spoiled
and all these quantities would decrease with the temperature.
\begin{figure}[h!]
\begin{center}
\includegraphics[width=0.6\textwidth]{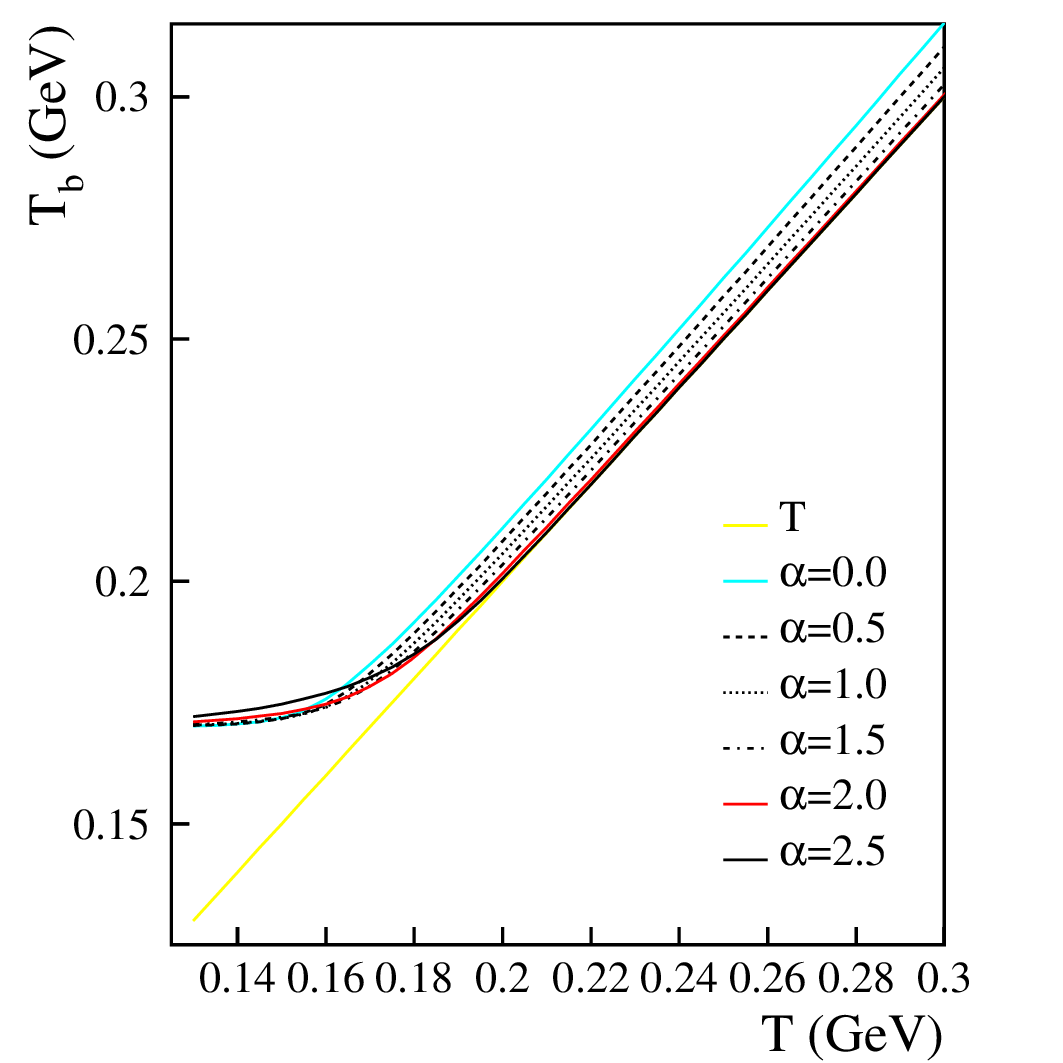}
\caption{\small{(Color online) The average effective bag temperature. 
The $45^\circ$'s straight line corresponds to the system temperature $T$.}}
\label{tb}
\end{center}
\end{figure}

It is also interesting to plot the average bag effective ``temperature'' $T_b\equiv k(p_k+B)^{1/4}$. 
As shown in Fig.~(\ref{tb}), for $\alpha=2$ and $\alpha=5/2$ this converges to $T$
very quickly above $T_0$. For $\alpha=3/2$ or smaller the effective bag temperature is always 
larger than the system temperature. 
This fact is a direct consequence of the inequality in Eq.~(\ref{presslim}) that 
gives
\begin{equation}
T_b = k(p_k+B)^{1/4} > T \; ,
\label{tlarger}
\end{equation}
Notice also that in the region $T<T_0$, $T_b\approxeq T_0$ for any value of $\alpha$.
In other words, the pressure $p_k$ is negligible with respect to $B$ 
and the system behaves as a standard hadron gas. 
It is worth mentioning that in our framework the compressible hadrons can exchange energy
only through a mechanical work (compression). They are, therefore, 
thermally insulated from the rest of the system and
the bag temperature, as well as the entropy, must be understood as 
quantities that measure the degeneracy of the hadronic states. 

\begin{figure}[h!]
\begin{center}
\includegraphics[width=0.9\textwidth]{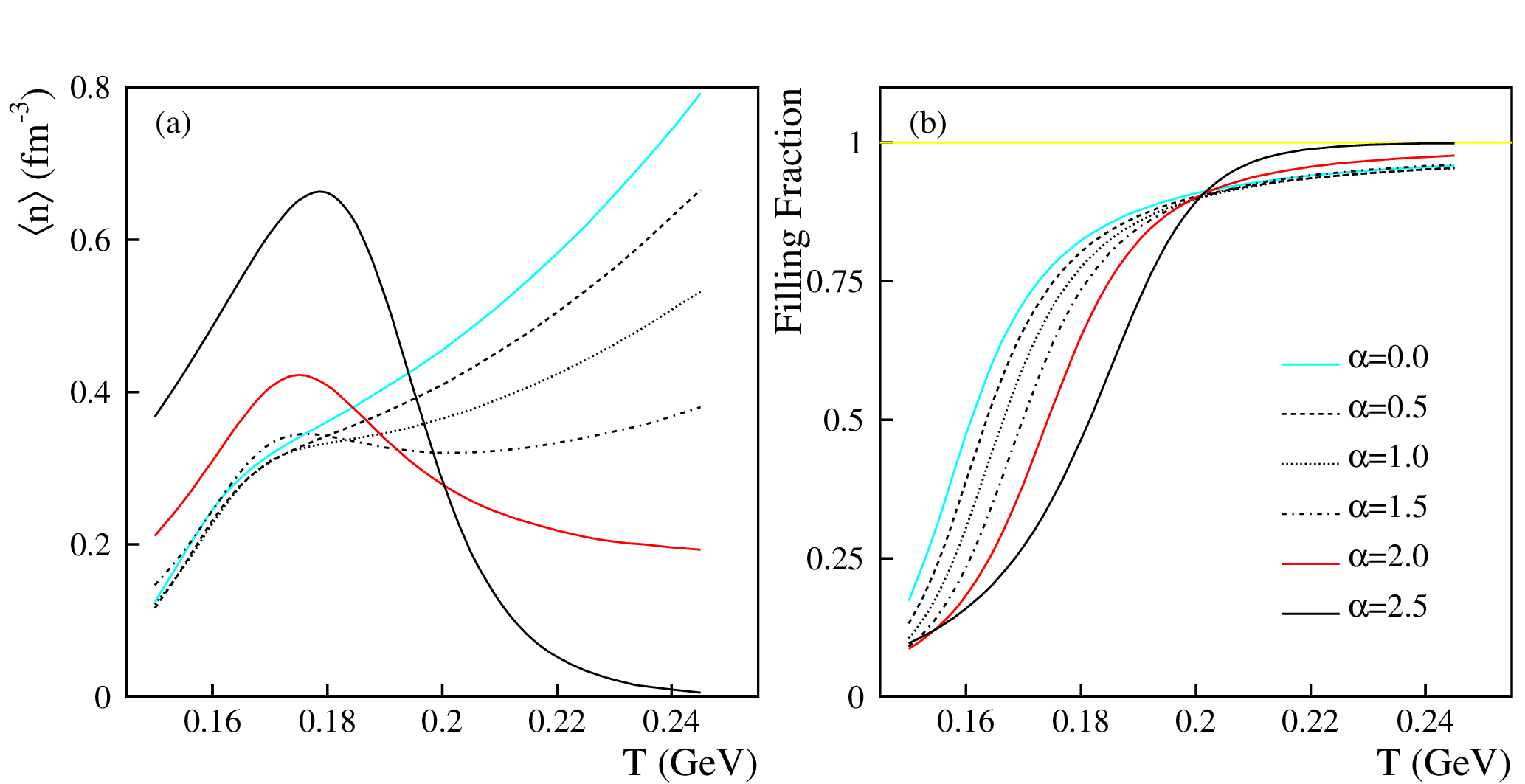}
\caption{\small{(Color online) Left panel (a): 
average number of particles per unit volume.
Right panel (b): The filling fraction ($f.f.$). The horizontal line $f.f.=1$ has been drawn for reference.}}
\label{new1}
\end{center}
\end{figure}

Finally, for $1 \leq \alpha \leq 5/2$, our model seems to produce a smooth crossover transition 
toward a new regime whose features are very similar to those of a gas of massless particles, 
even though no deconfined states are included in the partition function. 
To better understand this behavior it is useful to study the 
particles density $\langle n \rangle \equiv \langle N \rangle /V$ (Fig.~(\ref{new1}a))
and the {\em filling fraction} ($f.f.$) (Fig.~(\ref{new1}b)) which is defined as:
\begin{equation}
f.f. \equiv \frac{\langle V_{\rm hadrons} \rangle}{V}\; ,
\label{3.5a}
\end{equation}
where $\langle V_{\rm hadrons} \rangle$ is the average volume 
occupied by the hadrons (for a rigorous definition and formula see Appendix B).
The particles density can be calculated from the isobaric partition function
by introducing a fictitious fugacity $\lambda$ (to be set to 1 afterward) 
for each particle in the system, i.e. replacing $Z(V,T)$ with  
\begin{equation}
Z(V,T,\lambda) \equiv 
\sum_{N=0}^\infty \lambda^N Z_N (V,T) \; .
\label{fugaz}
\end{equation}
Accordingly, the isobaric partition function in Eq.~(\ref{ibpf2}) becomes
\begin{equation}
\widehat{Z}(T,s,\lambda)=\frac{1}{s-\lambda f(T,s)}
\label{fugaz2}
\end{equation}
and the corresponding solution for the pressure $p_k\equiv p_k(\infty,T,\lambda)$.
In the infinite volume limit, the particles density can be then obtained as:
\begin{equation}
\langle n \rangle= \lim_{V \rightarrow \infty} \frac{1}{V}\left. \frac{\partial \ln Z(V,T,\lambda)}{\partial \lambda}\right|_{\lambda=1}=
\frac{1}{T}\left. \frac{\partial p_k(\infty,T,\lambda)}{\partial \lambda}\right|_{\lambda=1}
\label{fugaz3}
\end{equation}
where in the last equality we have used the known relation $\ln Z= p V / T$.
As shown in Fig.~(\ref{new1}a), the particles density 
grows very rapidly as $T$ approaches $T_0$ from below. This is qualitatively what one expects for the hadron gas, 
where the average number of particles shows a monotonically growing behavior. Conversely, for $T>T_0$
this quantity depends very strongly on the choice of the parameter $\alpha$.
For $\alpha=0$, $1/2$ and $1$ we observe a change in the slope, but the curves still grow monotonically.
For $\alpha=1.5$ and $2$ the particles density has a local minimum at $T \sim 0.2$~GeV and $T \sim 0.27$~GeV, 
respectively (the latter lies outside the plotted region), and then it starts growing again with smaller slopes for
larger values of $\alpha$. For $\alpha=5/2$ 
the local minimum has disappeared, and after a sharp maximum at $T\sim 0.18$~GeV the particles density goes to zero.
This is the effect of the finite size of hadrons, which tends to saturate
the available system volume. In fact, as shown in Fig.~(\ref{new1}b),
when $T < T_0$ the filling fraction, Eq.~(\ref{3.5a}), is relatively small, whereas for 
higher temperatures, the system is almost totally filled by extended particles, 
i.e. $\langle V_{\rm hadrons} \rangle \sim V$. 
In this scenario, the space and the phase space available is strongly suppressed, and the
system tends to be populated by a smaller number of heavy particles. This effect 
is strongest for $\alpha=5/2$ (the filling fraction converges very fast to 1) 
and therefore $\langle n \rangle \rightarrow 0$. For smaller values of $\alpha$ this saturation 
effect becomes slightly less pronounced. A closer inspection reveals that for $\alpha < 1$ 
the filling fraction has a
maximum at very high temperature and then decreases with a very small 
slope\footnote{For $\alpha=0$, the rate of decrease of the filling fraction for large temperatures 
is maximum, but still only $\sim 2\%$ 
going from $T=1$~GeV to $T=10$~GeV}, whereas for $\alpha=1$ it seems to settle to a constant 
value $f.f.\sim 0.98$. The phase space, therefore, is never entirely suppressed. 

The behavior of the system changes continuously by varying 
$\alpha$. A numerical analysis indicates that there exist a value $\alpha_0$ between
$2.12$ and $2.13$ such that, at high $T$, the particles density vanishes for any 
$\alpha_0 < \alpha \leq 2.5$. In this range, the system is populated by one or 
few infinitely extended hadrons that occupy 
the entire space, filling the system with their inner QGP matter, which 
is a possible scenario for the deconfined phase. 
Conversely, for $1 \leq \alpha \leq \alpha_0$, we find many, rather 
heavy,``squeezed'' hadrons, which nonetheless mimic an ideal gas of massless particles. 
The number of particles, however, might be affected by the introduction of a
surface energy term in the spectrum in Eq.~(\ref{1.7}). 
Such a term would result in an energy cost associated with the 
splitting of a large hadron into many smaller ones and
might then widen the range of values for $\alpha$ that lead to $\langle n \rangle \simeq 0$
at temperatures above $T_0$.    
  
Note that, even though $\langle n \rangle$ can have a minimum or even vanish, 
the entropy of the system always increases monotonically with the temperature 
(see Fig.~(\ref{isob2}b)).
This is due to the fact that the dominant contribution (already at $T \sim T_0$)
comes from the bags entropies. This has been checked 
by using the classical expression 
for a gas with $\langle N \rangle$ particles (with the excluded volume correction) and degeneracy 
$g$:
\begin{displaymath}
S\sim S_{\rm class}(\langle N \rangle, g) = S_{\rm class} (\langle N \rangle, g\equiv 1) + \langle N \rangle \ln g \; .
\end{displaymath}
The contribution $\langle N \rangle \ln g$ corresponds to the sum of the bag entropies in Eq.~(\ref{entropy}), 
and it is dominant at high $T$.

\begin{figure}[h!]
\begin{center}
\includegraphics[width=0.6\textwidth]{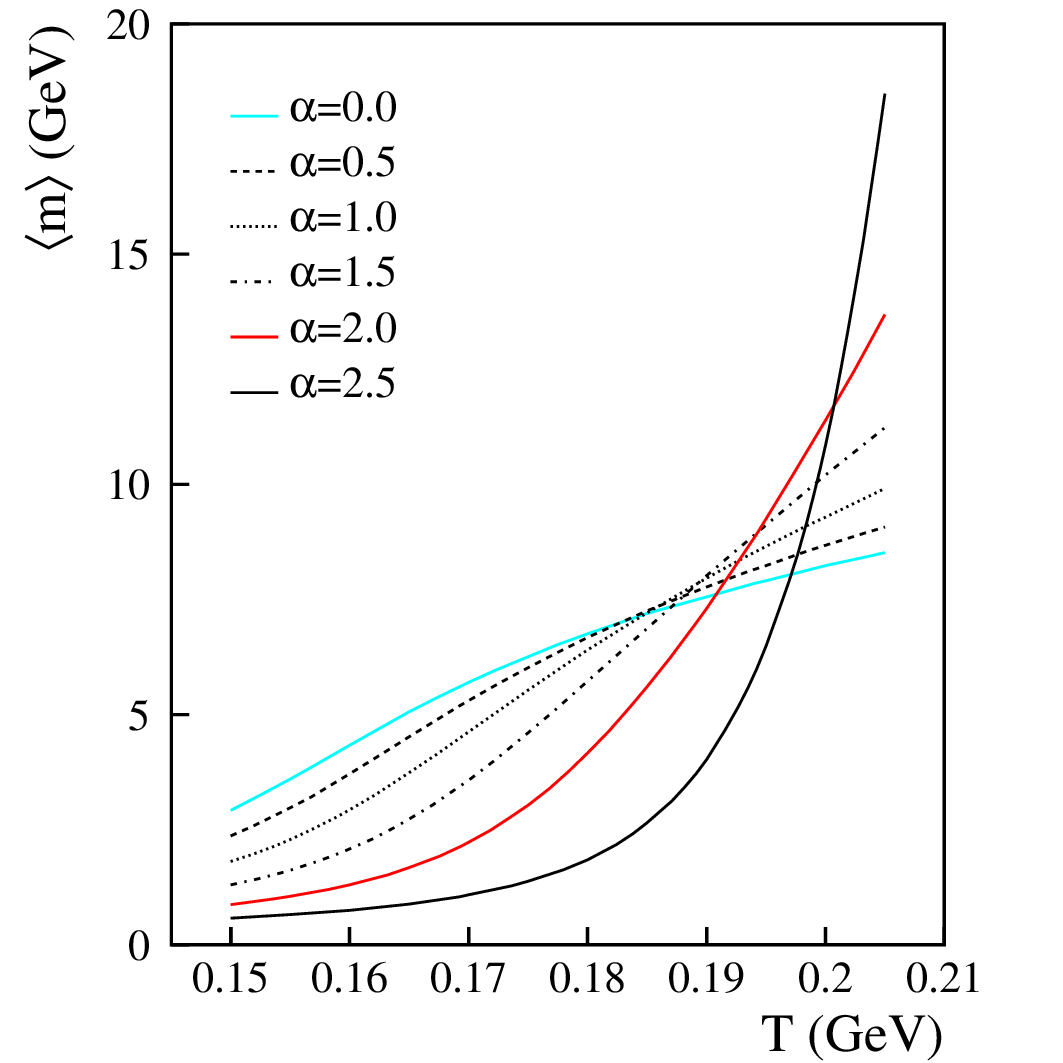}
\caption{\small{(Color online) Average hadrons mass.}}
\label{new2}
\end{center}
\end{figure}
Another interesting quantity is the average hadron mass $\langle m \rangle$, shown in Fig.~(\ref{new2})
that has been evaluated according to
\begin{equation}
 \langle m \rangle=  \frac{\varepsilon}{\langle n \rangle} - \frac{3}{2} T \;
\label{amass}
\end{equation}
As one can see, smaller values of $\alpha$ correspond to higher $\langle m \rangle$ at low temperature, 
whereas the situation is reversed at high $T$.
This follows from Eq.~(\ref{amass}) as a direct consequence of the 
behavior of $\langle n \rangle$ (Fig.~(\ref{new1}a)).
For $\alpha=5/2$ the average mass $\langle m \rangle$ grows with the temperature 
with the maximum slope and diverges when $\langle n \rangle \rightarrow 0$, i.e. when the system is populated 
by a finite number of infinite ``hadrons''. 

In the vicinity of the transition region $\langle m \rangle$ varies from $4.34$~GeV (for $\alpha=0$) to 
$0.75$~GeV (for $\alpha=5/2$) at $T=0.16$~GeV, 
and from $5.70$~GeV to $1.08$ at $T=0.17$~GeV, for $\alpha=0$ and $5/2$, respectively.
These values will somewhat depend on the choice of the model 
parameters. They would depend even more on an eventual upper mass cutoff.
In fact, the high value of $\langle m \rangle$ for $\alpha=0$ results from 
our assumption of an infinite mass spectrum and already at $T=0.16$~GeV 
it falls outside the region of the known hadrons. 
Our estimates are, however, lower than what one obtains for an exponential spectrum of
point-like hadrons as in Eq.~(\ref{1.7a}).
In such a case, for $\alpha=0$ the average mass is $\langle m \rangle=6.80$~GeV at $T=0.16$~GeV and 
infinity at $T=T_0=0.17$~GeV (the partition function itself is divergent). 
A mass cutoff at $2.0$~GeV would reduce these numbers to  $\langle m \rangle=1.36$~GeV 
and $\langle m \rangle=1.43$~GeV at  $T=0.16$~GeV and $T=0.17$~GeV, respectively.

\subsection{Consistency check}
\label{cons}

As a final remark, we want to discuss the consistency of the model.
To check this point we must make sure that quantities such as the energy density
(that we have evaluated from the pressure by using Eq.~(\ref{isobeps})) correspond 
to a thermal average of the form
\begin{equation}
\varepsilon =  \frac{1}{V}\frac{\sum_{\rm states}E_{\rm state}\e^{-E_{\rm state}/T}}
{\sum_{\rm states}\e^{-E_{\rm state}/T}} \;.
\label{concheck1}
\end{equation}  
A first hint in this direction, 
is given, {\em a posteriori}, by the results shown in this section, 
particularly, by quantities such as the filling fraction.
The $f.f.$ has been evaluated by making use of the relation in Eq.~(\ref{amass})
(see Appendix B) that implicitly relies on a form like Eq.~(\ref{concheck1}) for the energy density. 
Indeed, a wrong thermodynamical interpretation of $\varepsilon$ 
would likely have lead to dramatic consequences on the filling fraction, which contrarily
assumes only physical values in the interval $[0,1]$. 
However, to make a more direct test, we will provide an approximate expression for the 
grand-canonical partition function and we will compare $\varepsilon$ in Fig.~(\ref{isob1}b) 
with its corresponding value obtained as in Eq.~(\ref{concheck1}). 
This can be done for the case $\alpha=3/2$. For this particular value of $\alpha$ the multiple 
integrals over $\{ \d \eta_i \}$ in Eq.~(\ref{partfunmpr}) can be reduced to an unidimensional 
integral (see Appendix C) that greatly facilitates the numerical treatment.
Although, as demonstrated in this section, results do depend on the choice of $\alpha$, for the range of 
$0 \leq \alpha \leq 5/2$, one can hope that the following arguments will be still valid. 

Using $\alpha = 3/2$ in Eq.~(\ref{partfunmpr}) we get for the partition function
\begin{equation}
Z(V,T)=  \sum_{N=0}^{\infty} Z_N(V,T)
\end{equation}
with
\begin{eqnarray}
\label{gcpartfun}
&&Z_N(V,T) = 
\left[ c_0 \left( \frac{T}{2 \pi}\right)^{3/2}\right]^N \frac{1}{N!}  
\left[ \prod_{i=1}^N \int_{m_c}^{\infty} \d \eta_i 
\right]\nonumber \\  \; 
&&\times 
\exp{\left\{\left[ \frac{1}{kp_{r,N}^{1/4}}-\frac{1}{T}\left(\frac{3}{4}+\frac{B}{4p_{r,N}}\right)
\right]\sum_{i=1}^N \eta_i\right\}}\\ \nonumber \; 
&&\times 
\left(V-\sum_{i=1}^N \frac{\eta_i}{4p_{r,N}} \right)^N \Theta{\left( V-\sum_{i=1}^N\frac{\eta_i}{4p_{r,N}}  
\right)}\; .
\end{eqnarray}
Here we have added the suffix $N$ to $p_r$ to indicate the dependence of the pressure $p_{r,N}$ on 
the particle number in the canonical ensemble. 
The pressure $p_{r,N}$ is now just a parameter of the 
model. 
To determine its value, we need to find the value of $p_{r,N}$ that maximizes the logarithm of the 
integrand in Eq.~(\ref{gcpartfun}):
\begin{equation}
\Phi = \left[ \frac{1}{kp_{r,N}^{1/4}}-\frac{1}{T}\left(\frac{3}{4}+\frac{B}{4p_{r,N}}\right)
\right]\sum_{i=1}^N \eta_i+N \log
\left(V-\sum_{i=1}^N \frac{\eta_i}{4p_{r,N}} \right) \; ,
\label{integrand}
\end{equation}
where (for the moment) we omit the $\Theta$ function.  
The above expression depends on $\mathscr{M}\equiv \sum_{i=1}^N \eta_i$. To 
simplify the following derivation, we will introduce an approximation. 
Instead of solving $\partial \Phi / \partial p_{r,N} = 0$ 
for a generic set $\eta_1,\ldots,\eta_N$, we find a solution for the most important configurations 
defined by the value $\tilde{\mathscr{M}}$ that maximizes the integrand. In other words, 
we solve the system of equations:
\begin{equation}
\left\{
 \begin{array}{rl}
  &\frac{\partial \Phi}{\partial \mathscr{M}} = 0 \\
  &\frac{\partial \Phi}{\partial p_{r,N}} = 0
 \end{array} \right. \;.
\label{system}
\end{equation}
Note that the second condition ensures that $\varepsilon=T^2(\partial \ln Z / \partial T)/V$ 
has the form of Eq.~(\ref{concheck1}) as any implicit dependence on $T$ of $p_{r,N}$
do not contribute to $\partial \ln Z / \partial T$. 
In fact, if we denote with $\partial^*\ln Z /\partial T$ the derivative performed only 
on the explicit $T$ dependence of $\ln Z$ we have
\begin{equation}
\frac{\partial \ln Z}{\partial T} = \frac{\partial^* \ln Z}{\partial T} + \frac{1}{Z}
\sum_{N=0}^{\infty} \frac{\partial  Z_N}{\partial p_{r,N}} \frac{\partial p_{r,N}}{\partial T} 
\label{addit}
\end{equation}
where the second term vanishes because of the second condition in Eq.~(\ref{system}).
The Eq.~(\ref{system}) results in (see Appendix D):
\begin{equation}
\tilde{p}_{r,N}=B+\frac{N T}{\left(V-\frac{\tilde{\mathscr{M}}}{4\tilde{p}_{r,N}}\right)} \; .
\label{cccheck}
\end{equation}
The interpretation of the above expression is straightforward. Writing 
$\tilde{p}_{r,N} \equiv B+\tilde{p}_{k,N}$ one obtains the equivalent equation:
\begin{equation}
\tilde{p}_{k,N} \equiv \frac{N T}{V- \frac{\tilde{\mathscr{M}}}{4(B+\tilde{p}_{k,N})}} \;. 
\label{eq2a}
\end{equation}
Here, the right-hand side of Eq.~(\ref{eq2a}) is simply the 
canonical pressure of an ideal gas of $N$ particles with the total volume $V$  
replaced by the available volume $(V-\sum_i^N V_i)$. It is then natural to identify $p_{k,N}$ 
with the kinetic pressure in the canonical ensemble. Of course, once averaged 
over $N$ and over $\{\eta_i\}$, for a sufficiently large $V$, this
pressure must coincide with $p_k(\infty,T)$ evaluated with the isobaric partition function.
Eq.~(\ref{eq2a}) has the form 
of a self-consistency relation, as $\tilde{p}_{k,N}$ appears also in the right-hand side in the excluded volume
term.
Eq.~(\ref{eq2a}) is a quadratic form and has two solutions: a negative and positive one, 
and the negative corresponds to the situation where volume of hadrons exceeds 
the total volume, $\sum_{i=1}^N V_i > V$. The positive solution on the other hand ensures 
$ \sum_{i=1}^N V_i \leq V$ and therefore the condition for the $\Theta$ function in Eq.~(\ref{gcpartfun}) 
is always fulfilled.
In what follows, we will adopt the positive solution of the Eq.~(\ref{eq2a}) for any $\{\eta_1,\ldots,\eta_N \}$ 
(not only for the most probable set $\sum_i^N \eta_i=\tilde{\mathscr{M}}$)
and we will integrate numerically $Z_N(V,T)$ in the variables $\{ \eta_i \}$. 
The partition function $Z(V,T)$ is then evaluated by summing over the particle number 
$N$ up to a cutoff $N_{cut}$
\begin{equation}
Z(V,T)=\sum_{N=0}^{N_{cut}} Z_N(V,T)
\label{sumn}
\end{equation} 
where $N_{cut}$ is sufficiently large to ensure the accuracy of our calculations.
Finally we test the consistency of our picture by comparing 
the ratios $p_k/T^4$ and $\varepsilon/T^4$ with the results 
obtained with the isobaric partition function.
\begin{figure}[h!]
\begin{center}
\includegraphics[width=0.9\textwidth]{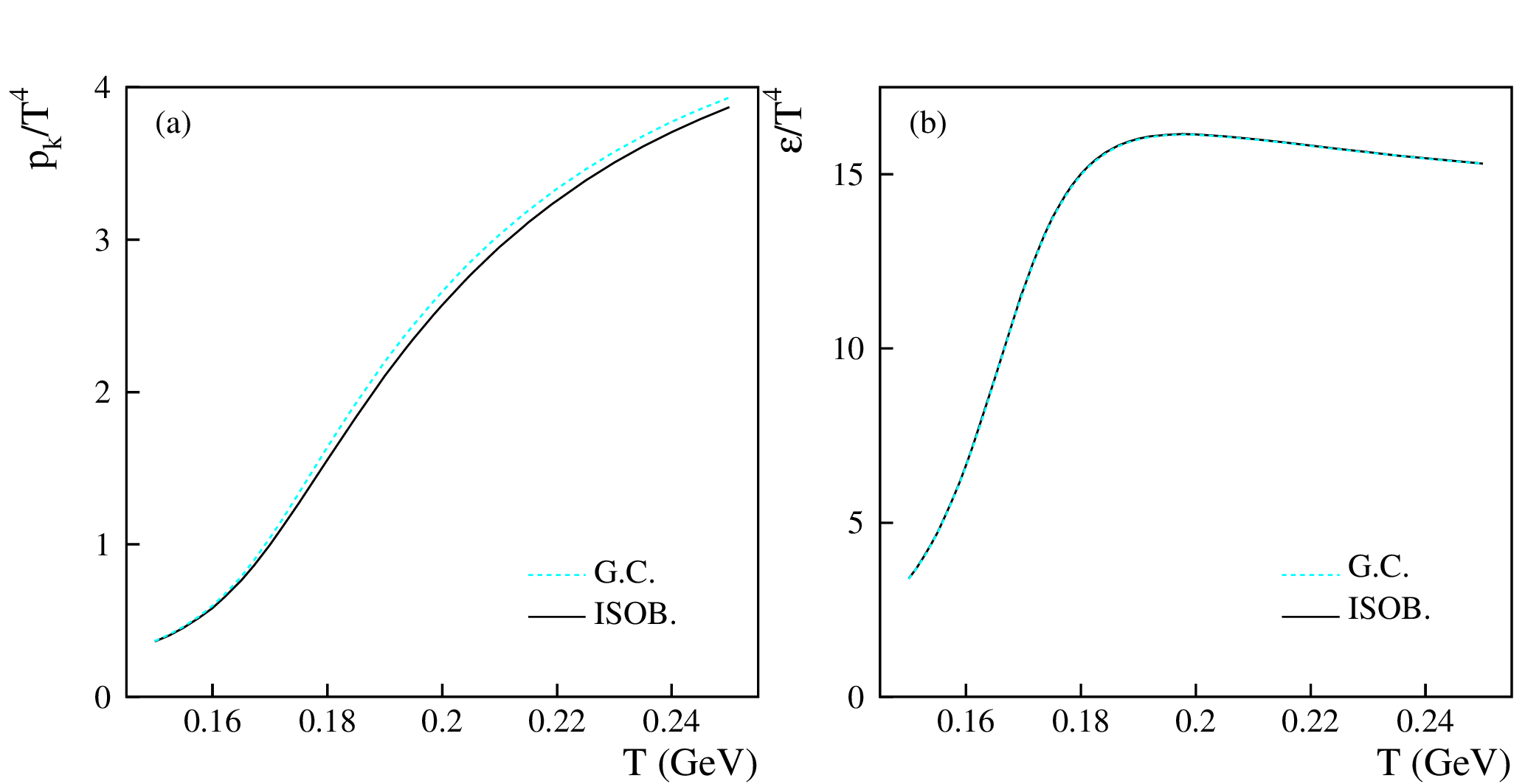}
\caption{\small{(Color online) Left panel (a): Comparison between the ratio $p_k/T^4$ evaluated with the 
isobaric partition function (solid
line) and
with the grand canonical partition function (dashed line).
Right panel (b): The same as in the left panel for the corresponding ratio $\varepsilon/T^4$.}}
\label{check}
\end{center}
\end{figure}
In Fig.~(\ref{check}), the ratios $p_k/T^4$ (left panel) and 
$\varepsilon/T^4$ (right) have been evaluated with the isobaric partition function
(solid line) and with the grand canonical partition function (dashed line) for a volume
$V=6.4\cdot10^4$~GeV$^{-3}$, which is, as we checked, a good approximation of the infinite volume limit.
As one can see, in both cases the two curves are very close, actually for the energy density they are 
practically coincident. The small (expected) difference between the isobaric and the grand-canonical result 
reflects the quality of our approximation.
The same test has been also performed on the entropy density, the particles density and the 
filling fraction. For all these quantities we have observed an equally good agreement.

\section{Conclusions and Discussion}
\label{conc}

\begin{figure}[h!]
\begin{center}
\includegraphics[width=1.0\textwidth]{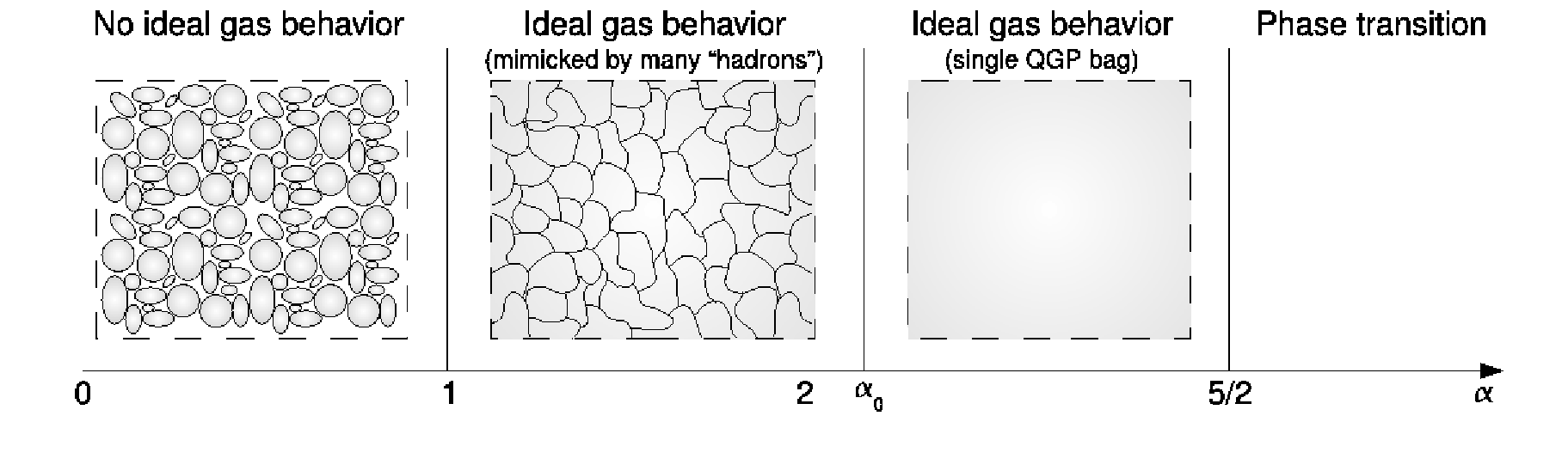}
\caption{\small{Pictorial representation of the various high-temperature phases of the model as a
function of $\alpha$. The scale has been distorted for visual reasons.}}
\label{schem}
\end{center}
\end{figure}

In this article, we have studied the crossover transition of the gas of bags~\cite{Gorenstein:1998am}.
We have found that the behavior of the system depends 
sensitively on the parameter $\alpha$ of the Hagedorn-like 
mass spectrum $\rho(m)=c_0m^{-\alpha}\exp{\{m/T_0\}}$. 
The system exhibits a crossover transition for $0 \leq \alpha \leq 5/2$ and an actual phase transition for
larger values.
In the range $0 \leq \alpha \leq 5/2$ we made a coarse scan of $\alpha$, setting $\alpha=0$, 
$1/2$, $1$, $3/2$, $2$ and $5/2$.
For $1 \leq \alpha\leq 5/2$ 
the gas of bags undergoes a sharp (yet, continuous) transition qualitatively
similar to lattice QCD. In this range, 
the asymptotic values of $p/T^4$, $\varepsilon/T^4$, $s/T^3$, coincide with the Stefan Boltzmann limit
for $\alpha=2$ and $\alpha=5/2$, whereas they settle to slightly larger values for $\alpha=3/2$ and $1$. 
For $\alpha =0$ and $1/2$, 
these quantities grow indefinitely with the temperature with small, decreasing, 
slopes going from $\alpha=0$ to $\alpha=1/2$. 
We have also studied the (strong) dependence of the particles density 
$\langle n \rangle = \langle N \rangle/V$ (where $V \rightarrow \infty$) on $\alpha$. 
We have found that there exist a limiting value $\alpha_0$ between $2.12$ and $2.13$ such that for
$\alpha_0<\alpha \leq5/2$ the particles density vanishes at high temperature. 
The system is then populated by one (or few) infinite bag(s) that occupies the entire volume.
Conversely, for $\alpha < \alpha_0$, $\langle n \rangle$ grows with the temperature. 
In the range $1 \leq \alpha < \alpha_0$, the ideal gas behavior is mimicked
by a number of heavy extended bags that saturates the phase space forming a dense system.
A pictorial representation summarizing the various high-temperature phases of the model is given in
Fig.~(\ref{schem}).

In this work we have explored a simple, intuitive model for a gas of hadrons 
in the vicinity of the transition at vanishing baryochemical potential.
To this end, we have adopted the idea of the MIT bag model and we have described 
hadrons as extended QGP bags. We have shown that, in the vicinity of $T_0$ 
(which is directly related to the transition temperature of our model), 
elastic interactions among hadrons play a fundamental role. In our schematic approach, 
they are quantified by the thermal pressure $p_k$. The effect of $p_k$ is to squeeze the hadrons, 
and for a certain set of model parameters, $1 \leq \alpha \leq 5/2$, the ideal gas behavior at high 
temperature can be reproduced. 
At the same time, the effective inner ``temperature'', or rather degeneracy parameter, $T_b$, 
of the bags increases with $p_k$, resulting in a temperature-dependent mass spectrum.
Above $T_0$, the physical picture of the QGP phase corresponds to a number $\sim 1$
of infinite bags that occupy the entire space. 
This is indeed the situation for $\alpha_0<\alpha \leq5/2$. 
On the other hand, a large number of independent QGP bags (such as for $\alpha < \alpha_0$)
would contradict the lattice findings of vanishing flavor-flavor 
correlations~\cite{Koch:2005vg,Allton:2005gk,Gavai:2005yk}.
However, in order to define precisely the range of values of $\alpha$ that lead
to a consistent QGP scenario, it is fundamental to study the effect of a 
surface energy term.
This contribution disfavors configurations with a large 
number of particles (they are more ``expensive'' in terms of surface energy) and 
might reduce the number of bags in the high-temperature phase, widening 
the range of possible values for $\alpha$.

Future work will likely concentrate on the effect of surface energy terms in addition 
to the study of van der Waals type residual interactions. 
It would be also interesting to analyze in detail the 
behavior of the system at finite baryochemical potential. This could be done 
starting from the formalism developed in~\cite{Gorenstein:2005rc}.

\section{Acknowledgments:}
The authors thank M.~I.~Gorenstein for a critical reading 
of the manuscript and for useful suggestions.
This work is supported  by the Director, Office of Energy
Research, Office of High Energy and Nuclear Physics, Divisions of
Nuclear Physics, of the U.S. Department of Energy under Contract No.
DE-AC02-05CH11231.

\section*{Appendix A}

By using the equations (\ref{simplass}) and (\ref{pkst}), the isobaric partition function in 
Eq.~(\ref{ibpf1}) can be conveniently 
written as:
\begin{eqnarray}
\label{appa1}
&&\widehat{Z}(T,s) = \sum_{N=0}^{\infty} \frac{1}{N!} \left[ \prod_{i=1}^N \int_{m_c}^{\infty} 
\d \eta_i\right] h_N(\{\eta_i \},T,s) \\ \nonumber \; 
&&\times
\int_{0}^{\infty}\d V \exp{\left[-sV\right]} \left(V-\sum_{i=1}^N \frac{\eta_i}{4 (B+sT)} \right)^N 
\Theta{\left( V-\sum_{i=1}^N\frac{\eta_i}{4(B+sT)}  
\right)} 
\end{eqnarray}
where
\begin{eqnarray}
\label{appa2}
&&h_N(\{\eta_i \},T,s)\equiv \left(\frac{T}{2 \pi}\right)^{3N/2}c_0^{N}
\left[ \prod_{i=1}^N 
\left(\frac{3}{4} \eta_i+ B \frac{\eta_i}{4(B+sT)} \right)^{3/2-\alpha} 
\right] \nonumber\\  \; 
&&\times
\exp{\left\{\left[ \frac{1}{k(B+sT)^{1/4}}-\frac{1}{T}\left(\frac{3}{4}+\frac{B}{4(B+sT)}\right)
\right]\sum_{i=1}^N \eta_i\right\}} \;.
\end{eqnarray}
The integral over $\d V$ can be carried out
\begin{eqnarray}
\label{appa3}
&&\int_{0}^{\infty}\d V \exp{\left[-sV\right]} \left(V-\sum_{i=1}^N \frac{\eta_i}{4 (B+sT)} \right)^N \Theta{\left( V-\sum_{i=1}^N\frac{\eta_i}{4(B+sT)}  
\right)} \\ \nonumber 
&&=\frac{N!}{s^{N+1}} \exp{\left\{ -\sum_{i=1}^N \eta_i \frac{s}{4(B+sT)} \right\}} \; .
\end{eqnarray}
and Eq.~(\ref{appa1}) becomes:
\begin{equation}
\label{appa4}
\widehat{Z}(T,s) = \frac{1}{s}\sum_{N=0}^{\infty}\left[ \prod_{i=1}^N \int_{m_c}^{\infty} 
\d \eta_i\right] \frac{h_N(\{\eta_i \},T,s)}{s^N}
 \exp{\left\{ -\sum_{i=1}^N \eta_i \frac{s}{4(B+sT)} \right\}} \; . 
\end{equation}
The multiple integral in the last equation, can be factorized as:
\begin{equation}
\label{appa5}
\left[ \prod_{i=1}^N \int_{m_c}^{\infty} 
\d \eta_i\right] h_N(\{\eta_i \},T,s)
 \exp{\left\{ -\sum_{i=1}^N \eta_i \frac{s}{4(B+sT)} \right\}} \equiv f(T,s)^N\; .  
\end{equation}
One then obtains Eq.~(\ref{ibpf2}) with $f(T,s)$ given by Eq.~(\ref{ibpf3a}).

\section*{Appendix B}

By using Eq.~(\ref{replacements}), we write the average volume occupied by hadrons as
\begin{equation}
\langle V_{\rm hadrons} \rangle= \langle N \rangle \frac{\langle U \rangle}{3 p_r}=
\langle N \rangle \frac{\langle m \rangle}{3p_r+B}=
\langle N \rangle \frac{\langle m \rangle}{3p_k+4B} \; ,
\label{appc1}
\end{equation}
where we used the stability condition $p_r=p_k+B$.
In turn, the average mass $\langle m \rangle$ reads (see also Eq.~(\ref{amass})):
\begin{equation}
\langle m \rangle=  \frac{\varepsilon}{\langle n \rangle} - \frac{3}{2} T \; .
\label{appc2}
\end{equation}
The filling fraction $f.f.$ can then be expressed in terms of $\varepsilon$, $p_k$ and the particles density 
$\langle n \rangle$ as:
\begin{equation}
f.f\equiv \frac{\langle V_{\rm hadrons} \rangle}{V}=
\frac{\varepsilon -3/2 \langle n \rangle T}{3p_k+4B} \; . 
\label{appc3}
\end{equation}

\section*{Appendix C}

The partition function of $N$ particles can be reduced to a unidimensional integral when
$\alpha=3/2$ as the factor $\left(3/4 \eta_i+ B V_i \right)^{3/2-\alpha}$ reduces to 1.
Let us begin by making the substitution: $\eta_i \rightarrow y^2_i+m_c$ that gives for
$Z_N$ in Eq.~(\ref{gcpartfun}):
\begin{eqnarray}
&&Z_N = \left[2 c_0
\left(\frac{T}{2 \pi}\right)^{3/2}\right]^{N}
\frac{1}{N!}
\left[ \prod_{i=1}^N \int_{0}^{\infty} \d y_i \; y_i
\right]\left(V-\frac{N m_c}{4 p_{r,N}}- \sum_{i=1}^N \frac{y^2_i}{4 p_{r,N}} \right)^N \nonumber \\  \; 
&&\times 
\exp{\left\{\left[ \frac{1}{kp_{r,N}^{1/4}}-\frac{1}{T}\left(\frac{3}{4}+\frac{B}{4p_{r,N}}\right)
\right]\left(\sum_{i=1}^N y^2_i+N m_c\right)\right\}}\; .
\label{a1}
\end{eqnarray}
It is now convenient to rewrite our integral by using $N$-dimensional hyperspherical coordinates by setting:
\begin{eqnarray}
\label{a2}
y_1&=& r \cos{\phi_1}\\ \nonumber
y_2&=& r \sin{\phi_1}\cos{\phi_2}\\ \nonumber
y_3&=& r \sin{\phi_1}\sin{\phi_2}\cos{\phi_3}\\ \nonumber
&\vdots&\\ \nonumber
y_N&=&r\sin{\phi_1}\sin{\phi_2}\ldots \sin{\phi_{N-1}} \; ,
\end{eqnarray}
and 
\begin{equation}
\label{a3}
\d^N r=r^{N-1} \sin^{N-2}{\phi_1}\sin^{N-3}{\phi_2} \ldots \sin{\phi_{N-2}} \; \d r \; \d \phi_1 \ldots \d
\phi_{N-1} \; .
\end{equation}
with
\begin{equation}
r^2\equiv \sum_i^N y^2_i \; .  
\end{equation}
We now note that, apart from $\d y_i \; y_i$, the integrand in Eq.~(\ref{a1}) depends only on
$r^2$. This is true also for the pressure $p_{r,N} = B+p_{k,N}$, where $p_{k,N}$ 
is given by the equations~(\ref{eq2a}) and, in the new variables, reads:
\begin{equation}
p_{k,N}=\frac{1}{6 V} \left( \xi+\sqrt{36 B\, T\, N\, V+\xi^2} \right)
\label{pkin}
\end{equation} 
where 
\begin{equation}
\xi=3TN+ \frac{3}{4}\left(r^2+N m_c \right)-3BV \; .
\label{a4}
\end{equation}
Eq.~(\ref{a1}) can now be written as:
\begin{eqnarray}
\label{a6}
&&Z_N = \left[2 c_0
\left(\frac{T}{2 \pi}\right)^{3/2}\right]^{N}\nonumber \\ \nonumber \; 
&&\times \frac{1}{N!}
\int_{0}^{\infty} \d r \; r^{N-1} \int_{0}^{\frac{\pi}{2}} \d \phi_1 \ldots \int_{0}^{\frac{\pi}{2}} \d \phi_{N-1}
\sin^{N-2}{\phi_1}\sin^{N-3}{\phi_2} \ldots \sin{\phi_{N-2}}\\ \nonumber \; 
&&\times
r^N \left[ \prod_{i=1}^{N-1} \cos{\phi_i} \right] \sin^{N-1}{\phi_1}\sin^{N-2}{\phi_2} \ldots \sin{\phi_{N-1}} 
\left(V-\frac{N m_c}{4 p_{r,N}}-  \frac{r^2}{4 p_{r,N}} \right)^N \nonumber \\  \; 
&&\times 
\exp{\left\{\left[ \frac{1}{kp_{r,N}^{1/4}}-\frac{1}{T}\left(\frac{3}{4}+\frac{B}{4p_{r,N}}\right)
\right]\left(r^2 +N m_c\right)\right\}}\; .
\end{eqnarray}
Next we separate Eq.~(\ref{a6}) into an angular integral, $A_N$, and a radial integral, $I_N$, such that:
\begin{equation}
Z_N = \left[2 c_0
\left(\frac{T}{2 \pi}\right)^{3/2}\right]^{N}\frac{1}{N!} A_N I_N 
\label{a7}
\end{equation}
where:
\begin{eqnarray}
\label{a8}
&&A_N = \int_{0}^{\frac{\pi}{2}} \d \phi_1 \ldots \int_{0}^{\frac{\pi}{2}} \d \phi_{N-1}
\sin^{N-2}{\phi_1}\sin^{N-3}{\phi_2} \ldots \sin{\phi_{N-2}}\\ \nonumber \; 
&&\times
 \left[ \prod_{i=1}^{N-1} \cos{\phi_i} \right] \sin^{N-1}{\phi_1}\sin^{N-2}{\phi_2} \ldots \sin{\phi_{N-1}} \; ,
\end{eqnarray}
and
\begin{eqnarray}
\label{a9}
&&I_N = 
\int_{0}^{\infty} \d r \;
r^{2N-1} \left(V-\frac{N m_c}{4 p_{r,N}}-  \frac{r^2}{4 p_{r,N}} \right)^N\\ \nonumber \; 
&&\times
\exp{\left\{\left[ \frac{1}{kp_{r,N}^{1/4}}-\frac{1}{T}\left(\frac{3}{4}+\frac{B}{4p_{r,N}}\right)
\right]\left(r^2 +N m_c\right)\right\}}\; .
\end{eqnarray}
The angular integral $A_N$ can be solved by applying the following recursion relation:
\begin{equation}
A_N=A_{N-1} \int_{0}^{\frac{\pi}{2}} \d \phi \sin^{2N-3}{\phi} \; \cos{\phi}=\frac{A_{N-1}}{2(N-1)} \; .
\label{a10}
\end{equation}
yielding
\begin{equation}
A_N=\frac{A_{N-1}}{2(N-1)}=\frac{A_{N-2}}{2^2(N-1)(N-2)}=\frac{A_{N-3}}{2^3(N-1)(N-2)(N-3)} \;.  
\ldots 
\label{a10b}
\end{equation}
Accordingly, by using $A_2=1/2$ one obtains:
\begin{equation}
A_N=\frac{1}{2^{N-1}(N-1)!} \; .
\label{a10a}
\end{equation}
The Eq.~(\ref{a6}) then reduces to:
\begin{eqnarray}
\label{a11}
&&Z_N = \left[2 c_0
\left(\frac{T}{2 \pi}\right)^{3/2}\right]^{N}\frac{1}{N!}\frac{1}{2^{N-1}(N-1)!} \int_{0}^{\infty} \d r \;
r^{2N-1} \\ \nonumber \; 
&&\times\left(V-\frac{N m_c}{4 p_{r,N}}-  \frac{r^2}{4 p_{r,N}} \right)^N
\exp{\left\{\left[ \frac{1}{kp_{r,N}^{1/4}}-\frac{1}{T}\left(\frac{3}{4}+\frac{B}{4p_{r,N}}\right)
\right]\left(r^2 +N m_c\right)\right\}}\; .
\end{eqnarray}

\section*{Appendix D}

Performing the derivatives, the system in Eq.~(\ref{system}) reads
\begin{equation}
\left\{
 \begin{array}{rl}
  & \frac{1}{kp_{r,N}^{1/4}} = \frac{N}{4 p_{r,N}}\frac{1}{\left(V- \frac{\mathscr{M}}{4p_{r,N}} \right)}+
\frac{1}{T}\left(\frac{3}{4}+\frac{B}{4 p_{r,N}}\right)\\
  & \frac{1}{kp_{r,N}^{1/4}} = \frac{B}{T p_{r,N}} + \frac{N}{p_{r,N}}\frac{1}{\left(V- \frac{\mathscr{M}}{4p_{r,N}} \right)}
 \end{array} \right. \;.
\label{systemapp}
\end{equation}
By subtracting the first equation from the second equation, one easily obtains Eq.~(\ref{cccheck}).


\end{document}